\documentclass[12pt]{article}
\usepackage{amsmath}
\usepackage{graphicx}
\usepackage{enumerate}
\usepackage{natbib}
\usepackage[colorlinks = true,
linkcolor = red,
urlcolor  = blue,
citecolor = blue,
anchorcolor = blue]{hyperref}
\usepackage{url} 
\usepackage{longtable}
\newcommand{\blind}{1}

\usepackage{comment}
\usepackage{xcolor}
\DeclareUnicodeCharacter{2212}{\ensuremath{-}}
\DeclareFontFamily{U}{mathx}{}
\DeclareFontShape{U}{mathx}{m}{n}{<-> mathx10}{}
\DeclareSymbolFont{mathx}{U}{mathx}{m}{n}
\DeclareMathAccent{\widehat}{0}{mathx}{"70}
\DeclareMathAccent{\widecheck}{0}{mathx}{"71}

\DeclareMathOperator*{\argmin}{arg\,min}
\usepackage{caption}
\usepackage{graphicx,latexsym,amssymb,amsmath}
\usepackage{subcaption}
\usepackage{bbm, bm}

\def\R{{\mathbb{R}}}

\def\E{{\mathbb{E}}}

\newcommand{{\convp}}{{\buildrel p\over\longrightarrow}}

\newcommand{\vertiii}[1]{{\left\vert\kern-0.4ex\left\vert\kern-0.4ex\left\vert #1 
		\right\vert\kern-0.4ex\right\vert\kern-0.4ex\right\vert}}

\newcommand{{\Vs}}{{\cal V}}
\newcommand{{\Ps}}{{\cal P}}
\newcommand{{\Ss}}{{\cal S}}
\newcommand{{\Xs}}{{\cal X}}
\newcommand{{\Ls}}{{\cal L}}
\newcommand{{\Ns}}{{\cal N}}
\newcommand{{\Zs}}{{\cal Z}}
\newcommand{{\Fs}}{{\cal F}}

\newtheorem{Lemma}{Lemma}[section] 
\newtheorem{Assumption}{Assumption}
\newtheorem{Theorem}{Theorem}[section] 

\newtheorem{Corollary}{Corollary}
\newcommand{\proofend}{$\hfill\Box{~}$}

\newcommand{\tcr}{\textcolor{black}} 

\usepackage{tikz}
\usetikzlibrary{tikzmark}
\usepackage{xcolor,colortbl}
\usetikzlibrary{positioning, automata,arrows}
\usetikzlibrary{arrows.meta,shapes.geometric}
\usetikzlibrary{decorations.pathreplacing,calligraphy}

\allowdisplaybreaks
\usepackage{xr}
\begin{document}

	\def\spacingset#1{\renewcommand{\baselinestretch}%
		{#1}\small\normalsize} \spacingset{1}
	
	
	\if1\blind
	{
		\title{\bf Factor-augmented sparse MIDAS regressions with an application to nowcasting\footnote{Jonas Striaukas is grateful to his advisor Eric Ghysels for his guidance and friendship during Jonas' PhD studies. The authors also thank Andrii Babii, Luca Barbaglia, Ferre De Graeve, Catherine Doz, Geert Dhaene, Domenico Giannone, S\'{i}lvia Gon\c calves, Laurent Ferrara, Julia Koh, Massimiliano Marcellino, Peter Reusens, Boriss Siliverstovs, Marie Ternes, Wouter Van der Veken, Raf Wouters, seminar and conference participants at CBS, COMPSTAT 2023, CREST, Paris School of Economics, EcoSta 2023, the MacroFor seminar of the International Institute of Forecasters and Erasmus University Rotterdam for helpful comments. Jad Beyhum undertook part of this work while employed by CREST, ENSAI (Rennes).}}
		\author{Jad Beyhum\\\vspace{0.5em}
			Department of Economics, KU Leuven, Belgium\vspace{0.5em}\\
			Jonas Striaukas
			\hspace{.2cm}\\
			Department of Finance, Copenhagen Business School, Denmark}
		\maketitle
	} \fi
	\if0\blind
	{
		\bigskip
		\bigskip
		\bigskip
		\begin{center}
			{\LARGE\bf Factor-augmented sparse MIDAS regression with an application to nowcasting}
		\end{center}
		\medskip
	} \fi
	
	\bigskip
	\begin{abstract}\noindent This article investigates factor-augmented sparse MIDAS (Mixed Data Sampling) regressions for high-dimensional time series data, which may be observed at different frequencies. Our novel approach integrates sparse and dense dimensionality reduction techniques. We derive the convergence rate of our estimator under misspecification due to the MIDAS approximation error, $\tau$-mixing dependence, and polynomial tails. Our method's finite sample performance is assessed via Monte Carlo simulations. We apply the methodology to nowcasting U.S. GDP growth and demonstrate that it outperforms both sparse regression and standard factor-augmented regression during the COVID-19 pandemic. These findings indicate that the growth through this period was influenced by both idiosyncratic (sparse) and common (dense) shocks. \tcr{The approach is implemented in the \texttt{midasml} R package, available on CRAN.}
	\end{abstract}
	
	\noindent%
	{\it Keywords:}  factor models, high-dimensional data, mixed-frequency data, nowcasting, COVID-19 pandemic
	\vfill
	
	\newpage
	
	\spacingset{1.5}
	
	\section{Introduction}\label{sec.intro}
	
	In the current economic climate, where accurate nowcasting of macroeconomic variables such as U.S. GDP is critical for policymakers, addressing the challenges posed by high-dimensional mixed-frequency data has become a key econometric concern. Nowcasting involves predicting the current or near-future values of low-frequency outcome variables using high-frequency data—a task complicated by the need for effective variable selection and control over parameter proliferation. Motivated by the challenges of the data used in nowcasting, this paper introduces a novel estimation method specifically designed for high-dimensional mixed-frequency data. 
	
	High dimensionality—where the number of predictors is large and can even exceed the number of observations—poses challenges for traditional estimation methods. Two prominent strategies have emerged: factor-augmented regressions and sparse methods. Factor-augmented regressions, as introduced by \citet{stock1999forecasting, stock2002forecasting}, reduce dimensionality by extracting a small set of latent factors via principal component analysis (PCA), under the assumption of a dense factor structure. These approaches are widely used for their efficiency and interpretability, especially when most predictors are assumed to contribute meaningfully. In contrast, sparse methods like LASSO \citep{tibshirani1996regression, bickel2003simultaneous} focus on selecting a small subset of relevant variables, promoting parsimony. In mixed-frequency settings—where variables are observed at different intervals, such as weekly financial and monthly macroeconomic series—additional complexities arise. MIDAS regressions \citep{andreou2010regression} address this by applying weighting schemes that directly relate high-frequency predictors to low-frequency outcomes. \tcr{Recent work adapts sparse and factor-augmented regressions to this context: \cite{siliverstovs2017short,uematsu2019high} propose LASSO approaches using an UMIDAS scheme}, \citet{babii2022machine} use a sparse-group LASSO on MIDAS-weighted variables to exploit natural group structures, while \citet{marcellino2010factor}, \citet{andreou2013should}, and \citet{koh2023inference} examine factor MIDAS regressions, where factors estimated by PCA are aggregated via nonlinear MIDAS schemes.
	
	In this paper, we propose the novel approach of factor-augmented sparse MIDAS regression for forecasting with mixed-frequency high-dimensional data, which harnesses the advantages of both LASSO and factor-augmented regression. Our method works as follows. First, we construct MIDAS-weighted variables using pre-set polynomials, enabling the flexible handling of mixed-frequency data. Second, we estimate factors through PCA applied on the MIDAS-weighted variables. Third, we estimate a high-dimensional linear regression model using the sparse-group LASSO estimator in which we include the MIDAS-weighted covariates and the extracted factors as predictors. This integrated approach captures both dense signals (through the extracted factors) and the sparse signals distributed across different frequencies. By applying the sparse-group LASSO estimator, we effectively model hierarchical relationships among predictors, as it accommodates both grouped variables and sparsity within groups \citep{babii2022machine}. 
	
	We establish the theoretical properties of our estimator, deriving rates of convergence that account for potential misspecification due to the MIDAS approximation or approximate sparsity. Our theoretical framework accommodates $\tau$-mixing processes with polynomial tails, which are often encountered in financial and macroeconomic time series. This setting allows our method to remain robust even when facing heavy-tailed distributions or complex dependencies, as commonly observed in nowcasting applications. Note that it is important to use $\tau$-mixing—which was introduced by \cite{dedecker2004coupling}—as opposed to more traditionally used $\beta$- or $\alpha$-mixing in our setting. The reason is that both $\beta$ and $\alpha$ mixing coefficients are too strong for the process we consider, namely, an autoregressive distributed lag (ARDL) model, see \cite{babii2022machine} for a formal analysis of $\tau$-mixing processes within the context of the ARDL model. In the present paper, we extend the modeling approach \cite{babii2022machine} to a factor-augmented model and show its validity under the $\tau$-mixing assumption. Relative to \cite{babii2022machine}, the challenge is to obtain results when factors are estimated by PCA. 
\tcr{The theoretical results show that, when the number of variables used to estimate the factors is sufficiently large, factor estimation does not affect the convergence rate of the estimator. This reinforces the reliability of the method and strengthens the confidence with which researchers can apply it in practice.}

	Numerical results underscore the practical effectiveness of the proposed factor-augmented sparse MIDAS approach. Monte Carlo simulations show its superior finite-sample performance relative to existing methods. Empirically, we apply the method to nowcast U.S. GDP growth using panels of weekly financial and monthly macroeconomic data. By modeling macroeconomic predictors through a factor structure and integrating both sparse and dense signals, our approach outperforms sparse-only and standard factor-augmented regressions—particularly during the COVID-19 pandemic. These results suggest that large shock during pandemic was driven by both idiosyncratic (sparse) and common (dense) components. Importantly, we find that factor augmentation offers limited gains in stable periods, where sparsity alone suffices, but becomes essential during episodes of heightened macroeconomic uncertainty. This distinction highlights the value of our method in adapting to changing economic conditions and improving forecast accuracy when it matters most.
	
	\bigskip
	
	\noindent\textbf{Literature review.} Our theoretical results contribute to the literature on high-dimensional econometrics. \cite{babii2022machine} study the properties of the sparse-group LASSO with mixed-frequency data (without estimated factors). We extend their work by proving an estimation error bound and deriving convergence rates while allowing for estimated factors and $\tau$-mixing dependence. To obtain these results, we establish convergence rates for factor estimation using PCA under $\tau$-mixing, thereby providing novel insights into the statistical properties of PCA in time series settings. Considering $\tau$-mixing processes is important, as highlighted by \cite{dedecker2004coupling, babii2022machine} among others, as other commonly used dependence assumptions appear to be too strong for the models we consider.  We also want to stress the differences between the factor MIDAS method of \cite{marcellino2010factor,andreou2013should,ferrara2019nowcasting, koh2023inference} and our approach. Their approach assumes a factor model on a high-dimensional, high-frequency set of variables. They implement a two-step procedure: first, the latent factors are estimated via PCA, and then these factors are projected onto the target variable using a low-dimensional nonlinear MIDAS regression. Instead, our approach uses a high-dimensional linear MIDAS regression, which also allows the many original predictors to play a role as long as they exhibit a sparse coefficient pattern.

	Several recent studies have investigated high-dimensional factor-augmented sparse regression models outside the mixed-frequency context. \citet{fan2023bridging} propose a general framework that simultaneously accommodates sparse and dense signals. Building on this, \citet{beyhum2024testing} introduce a fully data-driven procedure to test for the presence of sparse components in such models, showing that sparsity is statistically significant for most FRED-MD series. \citet{krampe2021factor} propose a factor-augmented sparse VAR and demonstrate its superior forecasting performance. Sparse-plus-dense structures have also been explored in high-dimensional panel data settings by \citet{hansen2019factor} and \citet{vogt2022cce}, who incorporate factor structures alongside sparse idiosyncratic components. \tcr{Our work is the first to consider factor-augmented sparse regression in a mixed-frequency framework.} The previously mentioned papers do not consider MIDAS weighting, their theoretical results do not allow for approximation errors, and they rely on stronger dependence measures than $\tau$-mixing.

	
	\tcr{Alternative approaches have been proposed to go beyond purely sparse or factor-augmented regressions. A related strand of the literature on targeted predictors—such as \citet{bai2008forecasting} and \citet{bessec2013short}—adopts a two-step procedure: strong predictors are first selected using soft or hard thresholding, and factors are then extracted from the selected subset. This strategy emphasizes dense modeling conditional on pre-selection, in contrast to the integrated sparse-plus-dense framework we develop. \citet{franjic2024nowcasting} propose a sparse mixed-frequency dynamic factor model that directly combines sparsity with a dynamic factor structure, aiming to exploit both high-frequency signals and cross-sectional dependence. \citet{mogliani2021bayesian} cast the sparse MIDAS framework in a Bayesian setting using structured shrinkage priors, i.e., group LASSO type, while \citet{kohns2025flexible} consider time-varying coefficients with structured group-shrinkage priors.}

	Finally, our work contributes to the growing literature on nowcasting, which faced significant challenges during the COVID-19 pandemic due to unprecedented economic shocks. \citet{foroni2022forecasting} adjust nowcasts using prediction errors from the financial crisis, while \citet{huber2023nowcasting} employ non-parametric Bayesian regression trees for robust GDP forecasts. \citet{carriero2024addressing} address pandemic-related outliers using outlier-augmented stochastic volatility BVARs, and \citet{hauzenberger2024nowcasting} propose Bayesian MIDAS regressions with Gaussian Processes to capture nonlinear effects. Additionally, \citet{diebold2020covid} evaluates the effectiveness of the ADS index in tracking real-time economic activity during the pandemic. Our findings show that factor-augmented sparse methods, grounded in a rigorous statistical framework, offer a robust alternative for nowcasting in periods of heightened uncertainty. Empirically, we demonstrate that dense macroeconomic information becomes especially valuable during crises, yielding significantly improved predictions throughout the COVID-19 period.
	
	\bigskip

	\bigskip
	\noindent\textbf{Notation.} For an integer $N\in \mathbb{N}$, let $[N]=\{1,\dots, N\}$. The transpose of a $n_1 \times n_2$ matrix $H$ is written $H^{\top}$. Its $k^{th}$ singular value is $\sigma_k(H)$. Let us also define the Euclidean norm $\left\| H\right\|_2^2=\sum_{i=1}^{n_1} \sum_{j=1}^{n_2}H_{i,j}^2$ and the sup-norm $\|H\|_\infty=\max\limits_{i\in[n_1], j\in[n_2]}|H_{i,j}|$. The quantity $n_1 \vee n_2$ is the maximum of $n_1$ and $n_2$. For a real-valued random variable $\zeta$ and $g>0$, we let $\vertiii{\zeta}_g=\E[|\zeta|^g]^\frac1g$. For a $d$-dimensional random vector $\zeta$, we define $\vertiii{\zeta}_g=\sup\limits_{u\in\R^d: \ \|u\|_2\le 1}\vertiii{u^\top \zeta}_g$.   For a vector $v\in\R^N$, and $G\subset[N]$, we let $v_G$ be the vector in $\R^N$ such that $(v_G)_i=v_i$ if $i\in G$ and $(v_G)_i=0$ otherwise.

	\section{Factor-augmented sparse MIDAS regression}\label{sec:methods}
	
	In this section, we describe our novel factor-augmented sparse MIDAS regression. Let $\bigl\{y_t,\ t \in[T]\bigr\}$ be the low-frequency outcome variable we want to predict where $T$ denotes the sample size. For example, in our application, we consider U.S. real GDP growth, which is measured quarterly. We use information in two sets of high-frequency regressors denoted  $$\widetilde{z}=\left\{\widetilde{z}_{t-(j-1)/m_{z},k},\  t\in [T], k \in[K_{z}],j\in[m_{z}]\right\}$$ and $$\widetilde{x}=\left\{\widetilde{x}_{t-(j-1)/m_{x},k},\  t\in [T],k \in[K_{x}],j\in [m_{x}]\right\}.$$ Note that, throughout the paper, we use a ``$\sim$" to denote high-frequency variables. In total, there are $K=K_{z}+K_{x}$ regressors which we observe more frequently than the target variable $y_t$, e.g., weekly and monthly. The quantities $m_{z}$ and $m_{x}$ determine the number of high-frequency lags we use of variables in $\widetilde{z}$ or $\widetilde{x}$, respectively. In our application, $\widetilde{z}$ is a high-dimensional weekly panel of financial variables, while $\widetilde{x}$ is a high-dimensional monthly panel of macroeconomic predictors. We distinguish these data types because they differ in frequency, and because we assume that $\widetilde{x}$ follows a factor model whereas $\widetilde{z}$ does not.
	

	The nowcasting task of $y_t$ by $\widetilde{z}$ and $\widetilde{x}$ is a high-dimensional problem because (i) $\widetilde{z}$ and $\widetilde{x}$ contain many variables and (ii) the higher sampling frequency of the variables $\widetilde{z}$ and $\widetilde{x}$ creates parameter proliferation because many lags can be introduced in the model. This calls for the use of several complementary dimension reduction approaches.

	The model is
	\begin{equation}\label{model_reg}
		y_t = m_t +\varepsilon_{t}, \ t \in [T],
	\end{equation}
	where
	\begin{equation}\label{model_m}
		\begin{aligned}
			m_t&=\rho_0 + \sum_{j=1}^{J} \rho_j y_{t-j} 
			+ \sum_{k=1}^{K_z}\sum_{j=1}^{m_{z}}\omega_{z,k}\left((j-1)/m_{z}\right) \widetilde{z}_{t-(j-1)/m_{z},k}  \\ &+\sum_{k=1}^{K_x}\sum_{j=1}^{m_{x}}\omega_{x,k}\left((j-1)/m_{x}\right) \widetilde{x}_{t-(j-1)/m_{x},k} 
			+\sum_{r=1}^{K_f}\sum_{j=1}^{m_{x}}\omega_{f,k}\left((j-1)/m_{f}\right) \widetilde{f}_{t-(j-1)/m_{x},r},
		\end{aligned}
	\end{equation}
	is the true regression function. 
		Note that our theory does not rely on the definition of $m_t$ in \eqref{model_m} but only requires that the true $m_t$ is well approximated by our predictors. Hence, our theory implicitly allows the true model to contain nonlinearities. The definition of $m_t$ in \eqref{model_m} serves to justify our approach heuristically.
	Here, $\rho=(\rho_0,\dots,\rho_J)^\top\in\R^{J+1}$ are the autoregressive coefficients,  $J$ is the number of lags,  $\omega_{z,k},\ k\in[K_z]$, $\omega_{x,k},\ k\in [K_x]$ and $ \omega_{f,r},\ r\in[K_f]$ are some functions from $[0,1]$ to $\R$ determining the linear regression coefficients of the different high-frequency lags,  and $$\widetilde{f}=\left\{\widetilde{f}_{t-(j-1)/m_{x},r},\  t\in [T],r \in[K_{f}],j\in[m_{x}]\right\}$$ are factors linked to $\widetilde{x}$ through the factor model:
	\begin{equation}\label{fac_model}
		\widetilde{x}_{t-(j-1)/m_{x},k}=\sum_{r=1}^{K_f} \widetilde{b}_{k ,r}\widetilde{f}_{t-(j-1)/m_{x},r} +\widetilde{u}_{t-(j-1)/m_{x},k},  
	\end{equation}
	for $t\in [T],k\in[K_{x}],j\in[m_{x}],$ where $\left\{\widetilde{b}_{k ,r},\ k\in[K_{x}],r \in[K_{f}]\right\}$ are nonrandom loadings and $\left\{\widetilde{u}_{t-(j-1)/m_{x},k},\  t\in [T],k \in[K_{x}],j\in[m_{x}] \right\}$ are error terms.  Note that we only assume that $\widetilde{x}$ (and not $\widetilde{z}$) follows a factor model. This is because, following the literature, it is uncommon to assume a factor structure among financial variables.

	The model in \eqref{model_reg}-\eqref{model_m} contains both the original regressors and the factors.  Estimating all the coefficients of the high-frequency lags in \eqref{model_m} is a complex econometric task. To overcome this difficulty, we approximate the high-frequency lag polynomials through a MIDAS approach. Specifically, we consider a dictionary $(w_d)_{d\geq 0}$ of functions from $[0,1]$ to $\R$ for which there exist linear approximation coefficients $\left\{ \widetilde\alpha_{k,d},\ k\in[K_z], d\in[D]\right\}$,  $\left\{ \widetilde\beta_{k,d},\ k\in[K_x], d\in[D]\right\}$ and  $\left\{ \widetilde\gamma_{r,d},\ r\in[K_f], d\in[D]\right\}$ such that, for all $u\in[0,1]$, 
	\begin{equation}\label{midas_approx}
		\begin{aligned}
			&\sum_{d=1}^D \widetilde\alpha_{k,d}w_d(u) \approx \omega_{z,k}\left(u\right) ,\ k\in[K_z] ,\\
			&\sum_{d=1}^D  \widetilde\beta_{k,d}w_d(u) \approx \omega_{x,k}\left(u\right) ,\ k\in[K_x] ,\\
			&\sum_{d=1}^D  \widetilde\gamma_{r,d}w_d(u) \approx \omega_{f,r}\left(u\right) ,\ r\in[K_f] ,
		\end{aligned}
	\end{equation}
	where $D$ is the number of polynomials from the dictionary used to approximate high-frequency lag coefficients linearly. In practice, we use Legendre polynomials up to degree $3$ as approximating functions, which implies that $D=4$.

	For all $t\in[T]$, let us define the MIDAS-weighted variables 
	\begin{equation}\label{Aggregation}
		\begin{aligned}
			z_{t,D(k-1)+d}&=\sum_{j=1}^{m_{z}}w_d\left((j-1)/m_{z}\right) \widetilde{z}_{t-(j-1)/m_{z},k},\  k\in[K_z],d\in[D];\\
			x_{t,D(k-1)+d}&=\sum_{j=1}^{m_{x}}w_d\left((j-1)/m_{x}\right) \widetilde{x}_{t-(j-1)/m_{x},k},\  k\in[K_x],d\in[D];\\
			f_{t,D(r-1)+d}&=\sum_{j=1}^{m_{x}}w_d\left((j-1)/m_{x}\right) \widetilde{f}_{t-(j-1)/m_{x},r},\  r \in[K_f],d\in[D];\\
			u_{t,D(k-1)+d}&= \sum_{j=1}^{m_{x}}w_d\left((j-1)/m_{x}\right) \widetilde{u}_{t-(j-1)/m_{x},k},\  k\in[K_x],d\in[D].
		\end{aligned}
	\end{equation}
	Let $p_z=DK_z$, $p_x=DK_x$, $R=DK_f$, $z_{t}=(z_{t,1},\dots,z_{t,p_z})^\top$, $x_t=(x_{t,1},\dots, x_{t,p_x})^\top$, $u_t= (u_{t,1},\dots,u_{t,p_x})^\top$, $f_t=(f_{t,1},\dots,f_{t,R})^\top$.
	Due to \eqref{midas_approx}, we have 
	\begin{equation}\label{model_m_approx}
		y_t= \rho_0 + \sum_{j=1}^{\text{J}}\rho_j y_{t-j}  + z_{t}^\top\alpha +x_t^\top  \beta + f_{t}^\top\gamma +a_t +\varepsilon_t,
	\end{equation}
	where
	$$  a_t =m_t - \left(\rho_0 + \sum_{j=1}^{\text{J}}\rho_j y_{t-j}  + z_{t}^\top\alpha +x_t^\top  \beta + f_{t}^\top\gamma\right)\approx 0,$$
	is the approximation error and
	\begin{equation*}
		\begin{aligned}
			\alpha_{D(k-1)+d}&= \widetilde \alpha_{k,d}, k\in[K_z],d\in[D], \\
			\beta_{D(k-1)+d}&= \widetilde \beta_{k,d}, k\in[K_x],d\in[D], \\
			\gamma_{D(r-1)+d}&= \widetilde \gamma_{r,d}, r\in[K_f],d\in[D].
		\end{aligned}
	\end{equation*}
	are the linear regression coefficients. Note that no multicollinearity issue between $x_t$ and $f_t$ arises in model \eqref{model_m_approx}. This is because by \eqref{fm}, equation \eqref{model_m_approx} can be rewritten as $y_t= \rho_0 + \sum_{j=1}^{\text{J}}\rho_j y_{t-j}  + z_{t}^\top\alpha +u_t^\top  \beta + f_{t}^\top(\gamma +B^\top\beta) +a_t +\varepsilon_t,$ and $u_t$ and $f_t$ are not multicollinear.
	Remark also that
	that $x_t$ follows the factor model\begin{equation}\label{fm}\begin{aligned}
			x_{t}=Bf_t+u_t,\ t\in[T],
	\end{aligned}\end{equation}
	where $B=(b_1,\dots, b_{p_x})^\top$ is a $p_x\times R$ matrix, with $b_{D(k-1)+d}= (\widetilde b_{k,1},\dots,\widetilde{b}_{k,R})^\top,\ k\in[K_z],d\in[D]$.

	To present the estimation procedure, we rewrite \eqref{model_m_approx} in a matrix form. We introduce further notation. Let $w_t=(1,y_{t-1},\dots,y_{t-J}, z_t^\top,x_t^\top)^\top $, $W=(w_1,\dots,w_T)^\top$, $F=(f_1,\dots,f_T)^\top$, $A=(a_1,\dots,a_T)^\top$ and $\mathcal{E}  = (\varepsilon_{1}, \dots, \varepsilon_{T})^\top$. $W$ is a $T\times p$ matrix where $p=1+J+p_z+p_x$, $F$ is a $T\times R$ matrix, while $A$ and $\mathcal{E}$ are $T\times 1$ vectors. The  regression model \eqref{model_m_approx}, therefore, can be written in a matrix form
	\begin{equation}\label{eq:mat_model2}
		Y = W\delta +F\gamma + A+\mathcal{E},
	\end{equation}
	where $\delta=(\rho_0,\rho_1,\dots,\rho_J,\alpha^\top,\beta^\top)^\top.$
	To proceed with the estimation of \eqref{eq:mat_model2}, we estimate the factors $F$ by PCA. We let the columns of $\widehat{F}/\sqrt{T}$ be the eigenvectors corresponding to the leading $R$ eigenvalues of $XX^\top$. In practice, the number of factors $R$ is unknown and we replace it by an estimator $\widehat{R}$. In our empirical application to nowcasting, we use the growth ratio estimator of \cite{ahn2013eigenvalue},  but there exist various alternatives in the literature \citep[see][among others]{bai2002determining,onatski2010determining,bai2019rank,fan2022estimating}. Finally, we use the estimator
	\begin{equation}\label{eq:est}
		\begin{aligned}
			(\widehat{\delta},\widehat{\gamma})\in\argmin\limits_{d\in \R^{p}, c\in \R^{R}} \left\|Y-Wd- \widehat{F} c \right\|^2_2 + \lambda\Omega(d),
		\end{aligned}
	\end{equation}
	where $\lambda\in\R$ is the tuning parameter, and the sparse-group LASSO (sg-LASSO) penalty function is
	\begin{equation*}
		\Omega(d) = \mu \|d\|_1 + (1-\mu)\|d\|_{2,1},
	\end{equation*}
	which interpolates between the $\ell_1$ LASSO and group LASSO norms. Here, $\mu\in[0,1]$ is the interpolation parameter between the two norms. The latter is defined as $\|d\|_{2,1}= \sum_{G\in\mathcal{G}}\|d_G\|_2$. Such a penalty is beneficial in MIDAS regression settings because it encourages within-group sparsity, which allows for accurate weight function estimation, as well as across-group sparsity, which controls the dimensionality of regressors \citep{babii2022machine}. In practice, $\lambda$ and $\mu$ are selected via cross-validation. Specifically, we use 5-fold cross-validation, defining folds as adjacent blocks of sub-samples over the time dimension to take into account the time series dependence. We experimented with the number of folds and the overall conclusions remain the same. We stick with 5-fold cross-validation to match the sg-LASSO-MIDAS approach of \cite{babii2022machine}, which we benchmark against. For the nowcasting application, following \cite{babii2022machine}, we choose a group structure corresponding to the original high-frequency covariates, that is $\mathcal{G}=\{G_1,\dots,G_{1+K_z+K_x}\}$, where $G_1 =[J+1]$ and $G_\ell =\{J+1+(\ell-1)d:\ d\in[D]\} $ for $\ell\in\{2,\dots, 1+K_z+K_x\}$. This structure promotes sparsity across the original predictors.

	This is a sparse plus dense dimension reduction approach. Indeed, sparsity is induced through the penalty $\Omega(\cdot)$ and the presence of the factors, which are linear combinations of the variables in $X$, introduces a dense pattern in the estimation procedure. By combining three different dimension reduction approaches (MIDAS, factor models/PCA and LASSO), we propose a method appropriate for the challenges at hand.

	Lastly, we note that in our empirical application, we also explore factor-augmented and sparse MIDAS regressions as alternatives to our approach. Specifically, the factor-augmented MIDAS regression assumes that coefficients related to the sparse component are zero and employs OLS for estimation. Formally, this approach solves 
	\begin{equation}\label{eq:estfa}
		\begin{aligned}
			\min\limits_{c\in \R^{R}} \left\|Y- \widehat{F} c \right\|^2_2 ,
		\end{aligned}
	\end{equation}
	On the other hand, the sparse MIDAS regression imposes zero restrictions on $\gamma$ coefficients, for which we apply the sg-LASSO estimator: \begin{equation}\label{eq:ests}
		\begin{aligned}
			\min\limits_{d\in \R^{p}} \left\|Y- Wd \right\|^2_2 + \lambda\Omega(d),
		\end{aligned}
	\end{equation} These methods are referred to as FAMIDAS (equation \ref{eq:estfa}) and sg-LASSO-MIDAS (equation \ref{eq:ests}), respectively. Our main approach in equation \eqref{eq:est}, which uses both sparse and dense signals, is denoted as sg-LASSO-FAMIDAS.

	\section{Theory}\label{sec.th}
	
	Let us now investigate the theoretical properties of our approach. Our theoretical analysis allows for the approximation error $a_t$, which can be due to the MIDAS approximation, approximate sparsity or nonlinearites. Notably, we do not need to assume that the true model $m_t$ is linear or sparse (or takes the form \eqref{model_m}), but only that it is well approximated by a sparse linear combination of the predictors that we use. We consider an asymptotic regime where $T\to \infty$, and $p_x$ and $p_z$ go to infinity as a function of $T$. The number of factors is fixed with $T$. It would be possible to let it grow with $T$, but this would greatly complicate the theoretical analysis and, more importantly, the exposition of results.\footnote{See, e.g., \cite{beyhum2022factor,freeman2023linear} for an analysis with a growing number of factors.} The group structure $\mathcal{G}$ and the parameters $\delta$ and $\gamma$ are not random but can vary with $T$. For this theoretical analysis, $\mu$ is fixed with $T$. 
	We make the following assumptions.

	\begin{Assumption}\label{as.factors} It holds that
		\begin{enumerate}[\textup{(}i\textup{)}]
			\item\label{f4i} For all $t\in[T]$, $\E[f_tf_t^\top]=I_R$ and $B^\top B$ is diagonal;
			\item \label{f4ii}  All the eigenvalues of the $R\times R$ matrix $p_x^{-1} B^\top B$ are bounded away from $0$ and $\infty$ as $p_x\to \infty$;
			\item\label{f4iv} $\|B\|_\infty=O(1)$.
		\end{enumerate}
	\end{Assumption}

	\begin{Assumption}\label{as.moments}
		The following holds:
		\begin{enumerate}[\textup{(}i\textup{)}]  
			\item\label{taili} For all $t\in[T],k\in[p_x], \ell\in [p], r\in[R]$, it holds that \begin{align*}&\E[u_{t,k}]=\E[u_{t,k}f_{t,r}]=\E[u_{t,k}\varepsilon_t]=\E[f_{t,r}\varepsilon_t]=\E[w_{t,\ell}\varepsilon_t]=0;\end{align*}
			\item\label{tailii} There exist $q > 2$ and $C_1>0$, such that, for all $t\in[T]$, we have 
			$$\vertiii{u_{t}}_{2q}+    \vertiii{\varepsilon_{t}}_{2q}+   \vertiii{f_{t}}_{2q}+ \vertiii{w_{t}}_{2q}+\vertiii{p_x^{-1/2}\sum_{k\in[p_x]}u_{t,k}b_k}_{2q} \le C_1;$$
			\item\label{tailiii} There exists $C_2>0$ such that, for all $t\in[T]$,
			$\max\limits_{k\in [p]}\|E[w_{t,k}u_t]\|_2 \le C_2$ and
			$$\tcr{\max_{s,t\in[T]}\E\left[\left\{p_x^{-1/2}\left(u_s^\top u_t-\E[u_s^\top u_t]\right)\right\}^2\right]\le C_2.}$$
		\end{enumerate}
	\end{Assumption}
	
	Assumption \ref{as.factors} is the same as Assumption 3 in \cite{fan2023bridging}. Its conditions \eqref{f4i} and \eqref{f4ii} form a strong factor assumption as in \cite{bai2003inferential}. Assumption \ref{as.moments} contains restrictions on the moments of random variables. Its condition \eqref{taili} is a no-correlation condition. It would hold if (a) $u_t$ is mean zero and uncorrelated with $f_t,\varepsilon_t$ and (b) $\varepsilon_t$ is mean zero and uncorrelated with $f_t,w_t$. Condition \eqref{tailii} assumes that variables have strictly more than $4$ finite moments, having, therefore, polynomial tails. Finally, condition \eqref{tailiii} bounds some moments. \tcr{By the inequality of Cauchy-Schwarz, $\E\left[\left\{p_x^{-1/2}\left(u_s^\top u_t-\E[u_s^\top u_t]\right)\right\}^2\right]$ is bounded if $\max_{s,t\in[T]}\E\left[\left\{p_x^{-1/2}\left(u_s^\top u_t-\E[u_s^\top u_t]\right)\right\}^4\right]$ is bounded. A bound on the latter quantity is standard in the literature on factor models, see \cite{fan2013large,fan2023bridging}.} The bound on $\max\limits_{k\in [p]}\|E[w_{t,k}u_t]\|_2$ is new to the literature. It essentially assumes that $w_{t,k}$ can only be correlated with a few entries of $u_t$. 
	
	Next, to control the time series dependence of the variables, we use the concept of $\tau$-mixing. The definition of $\tau$-mixing coefficients is as follows. For a $\sigma$-algebra $\mathcal{M}$ and a random vector $\xi\in\R^l$, let
	\begin{equation*}
		\tau(\mathcal{M},\xi) = \bigg\|\sup_{f\in\mathrm{Lip}_1}\left|\E(f(\xi)|\mathcal{M}) - \E(f(\xi))\right|\bigg\|_1,
	\end{equation*}
	where $\mathrm{Lip}_1=\left\{f:\R^l\to\R:\;|f(x)-f(y)|\leq |x-y|_1\right\}$ is a set of $1$-Lipschitz functions. Let $\{\xi_t\}_t$ be a stochastic process and let $\mathcal{M}_t=\sigma(\xi_t,\xi_{t-1},\dots)$ be its filtration. The $\tau$-mixing coefficients of $\{\xi_t\}_t$ are defined as
	\begin{equation*}
		\tau_s = \sup_{j\geq 1}\frac{1}{j}\sup_{t+s\leq t_1<\dots<t_j}\tau(\mathcal{M}_t,(\xi_{t_1},\dots,\xi_{t_j})),\ s\in\mathbb{N}.
	\end{equation*}	
	The process $\{\xi_t\}_t$ is called \textit{$\tau$-mixing} when $\tau_s\downarrow0$ as $s\to\infty$ . As noted in the introduction, using $\tau$-mixing conditions is less restrictive than $\alpha$ or $\beta$-mixing. The latter are too strong for the process we consider, namely, an autoregressive distributed lag (ARDL) model, see \cite{dedecker2004coupling} and \cite{babii2022machine}.
	\begin{Assumption}\label{as.mixing}
		The process $$\left\{\left(f_{t}^\top, u_t^\top,\varepsilon_t, w_t^\top, \left(p_x^{-1/2}\sum_{k\in[p_x]}u_{t,k}b_k\right)^\top\right)^\top\right\}_t $$ is stationary, and, for some constants $C_3>0$ and $a>(q-1)/(q-2)$, its $\tau$-mixing coefficients, denoted by $\tau_s$ satisfy $\tau_s\le C_3s^{-a}$ for all $s\in\mathbb{N}$.
	\end{Assumption}
	
	We are ready to state some results on the estimation of the factors. To do so, we introduce further notation. Let $V$ be the matrix with the $R$ largest eigenvalues of $T^{-1}XX^\top$. Next, we define $H=T^{-1}V^{-1}\widehat{F}^\top FB^\top B$, which is a rotation matrix such that the factors rotated by this matrix are well estimated.\footnote{In the Online Appendix, we prove that $V$ is invertible with probability going to $1$, see Lemma \ref{OA-lm.V}.} This rotation matrix is the same as the one used in the literature to show consistency of estimated factors, see \cite{bai2003inferential} and \cite{bai2006confidence}. Moreover, we let $$h_T= \left(\frac{p}{T^{\kappa-1}}\right)^{1/\kappa}\vee \sqrt{\frac{\log(2p)}{T}},$$ where $\kappa= \frac{(a+1)q-1}{a+q-1} $. We have the following lemma.
	\begin{Lemma}\label{lm.fac.est} Under Assumptions \ref{as.factors}, \ref{as.moments} and \ref{as.mixing}, we have
		\begin{enumerate}[\textup{(}i\textup{)}]  
			\item\label{lfai} $\left\|\widehat{F}-FH^\top \right\|_2=O_P\left(\sqrt{\frac{T}{p_x}} +1\right)$;
			\item\label{lfaii} $\left\|\left(\widehat{F}-FH^\top\right)^\top \mathcal{E}\right\|_2=O_P\left(\sqrt{\frac{T}{p_x}} +1 \right);$
			\item\label{lfaiii} $\left\|\left(\widehat{F}-FH^\top\right)^\top W\right\|_\infty=O_P\left(\left(\frac{T}{p_x}+\sqrt{\frac{T}{p_x}}\right) \left(\sqrt{p_x}h_T+1\right)  \right) .$
		\end{enumerate}
	\end{Lemma}
	Lemma \ref{lm.fac.est} shows consistency of factor estimates. To the best of our knowledge, this is the first result of this type under $\tau$-mixing. The rates in statements \eqref{lfai} and \eqref{lfaii} are the same as in the standard literature, see \cite{bai2003inferential},  \cite{bai2006confidence} and \cite{fan2023bridging}. Hence the presence of $\tau$-mixing does not change the rate of convergence in $\ell_2$-norm. It only appears in the rate of convergence in $\ell_\infty$-norm such as statement \eqref{lfaiii}. 
	
	Now, let $\mathcal{S}_0=\{k\in[p]:\ \delta_k\ne 0\}$ and $\mathcal{G}_0=\{G\in\mathcal{G}:\ \delta_G\ne 0\}$ be respectively be the support and the group support of $\delta$. We define the effective sparsity $\sqrt{s_\mu}=\mu \sqrt{|\mathcal{S}_0|} +(1-\mu)\sqrt{|\mathcal{G}_0|}$ and the maximum group size $G^*=\max_{G\in\mathcal{G}}|G|$. To avoid notational complexities arising when $s_\mu=0$, we assume that $s_\mu \ge 1$ (otherwise, it suffices to replace $s_\mu$ by $s_\mu\vee 1$ in all statements). We also define $$\Sigma = \E[w_tw_t^\top]- \E[w_tf_t^\top] \E[w_tf_t^\top]^\top,$$
	which is the population Gram matrix of the $w_t$ once they have been projected on the vector space orthogonal to the factors $f_t$. We make the following assumption. 
	\begin{Assumption}\label{as.rate}
		The following holds:
		\begin{enumerate}[\textup{(}i\textup{)}]  
			\item\label{ratei}$\|\gamma\|_2=O(1)$, $p=o(T^{(\kappa-1)/2})$ and $$G^*s_\mu \left[\left(\left(\frac{p^2}{T^{\kappa-1}}\right)^{1/\kappa}\vee \sqrt{\frac{\log(2p)}{T}}\right)+\frac{1}{\sqrt{p_x}}\right]=o(1);$$
			\item\label{rateii} There exists $\nu>0$ independent of $T$ such that $\sigma_{p}(\Sigma)\ge \nu.$
		\end{enumerate}
	\end{Assumption}
	
	Assumption \ref{as.rate} \eqref{ratei} is a rate condition. It assumes that $p$ cannot grow too quickly with $T$, this is needed because of the presence of time series dependence and polynomial tails. It also restricts the rate at which the sparsity level $s_\mu$ can grow. In the literature on the LASSO estimator, it is typically assumed that $s_*\sqrt{\log(2p)/T}=o(1)$, where $s_*$ is the sparsity level \citep[see][]{bickel2003simultaneous}. Our condition is stronger because of time series dependence, polynomial tails and the factors are estimated. Condition \eqref{rateii} of Assumption \ref{as.rate} assumes that the smallest eigenvalue of $\Sigma$ is bounded from below. This allows to show that the classical restricted eigenvalue condition of the LASSO literature holds \citep{bickel2003simultaneous}.
	
	Then, we state a bound on the prediction error of our estimator, that is the $\ell_2$-norm of the prediction $W\widehat{\delta}+\widehat{F}\widehat{\gamma}$ minus the target prediction $ W\delta+F\gamma$. To state the bound, we need further notation. Let $P_{\widehat{F}}=\frac1T\widehat{F}\widehat{F}^\top$ be the orthogonal projector on the vector space spanned by the columns of $\widehat{F}$ and $M_{\widehat{F}}=I_T-P_{\widehat{F}}$. We also introduce $\widetilde{W} = M_{\widehat{F}}W $ which corresponds to $W$ projected on the orthogonal of the vector space spanned by the columns of $\widehat{F}$. Finally, let $\Omega^*(\cdot)$ be the dual norm of $\Omega(\cdot)$, that is, for any $d\in\R^p$, $\Omega^*(d)=\sup\limits_{v\in\R^p}\frac{v^\top d}{\Omega(d)}.$
	\begin{Theorem}\label{th.rate}  Under Assumptions \ref{as.factors}, \ref{as.moments}, \ref{as.mixing} and \ref{as.rate}, and letting $\lambda\ge 2\Omega^*\left(\frac1T\widetilde{W}^\top \mathcal{E}\right)$, with probability going to $1$, we have
		\begin{align*}&\frac{1}{\sqrt{T}}\left\|W\widehat{\delta}+\widehat{F}\widehat{\gamma}-W\delta-F\gamma\right\|_2\\
			&\le \left(\frac{288s_\mu\lambda^2}{\nu  }+ \frac{4}{T}\left(\|A\|_2+\left\|M_{\widehat{F}}F\gamma\right\|_2\right)^2\right)^{1/2} +\frac{1}{\sqrt{T}}\left(  \left\|M_{\widehat{F}}F\gamma\right\|_2+  \left\| P_{\widehat{F}} \mathcal{E}\right\|_2\right).
		\end{align*}
	\end{Theorem}
	As standard with estimation error bounds on the LASSO \citep{bickel2003simultaneous}, we need that the penalty term is large enough, that is  $\lambda\ge 2\Omega^*\left(\frac1T\widetilde{W}^\top \mathcal{E}\right)$. The quantity $\Omega^*\left(\frac1T\widetilde{W}^\top \mathcal{E}\right)$ is the effective noise of the problem \citep{lederer2021estimating}. In practice, $\lambda$ is chosen by cross-validation. As in the literature, the bound here depends on the approximation error $\|A\|_2$, see \cite{bickel2003simultaneous} and \cite{babii2022machine}. Compared to the traditional LASSO literature, our bound contains two additional terms:  $\left\|M_{\widehat{F}}F\gamma\right\|_2$ and  $\left\| P_{\widehat{F}} \mathcal{E}\right\|_2$. They are present because we do not use the true $F$ for estimation but rather its estimate $\widehat{F}$. The term $\left\|M_{\widehat{F}}F\gamma\right\|_2$ is the $\ell_2$-norm of the projection of $F\gamma$ on the orthogonal of the vector space generated by the columns of $\widehat{F}$, while the quantity $\left\| P_{\widehat{F}} \mathcal{E}\right\|_2$ is the $\ell_2$-norm of the projection of $\mathcal{E}$ on the vector space generated by the columns of $\widehat{F}$. Using in particular Lemma \ref{lm.fac.est}, we can bound $\left\|M_{\widehat{F}}F\gamma\right\|_2$,  $\left\| P_{\widehat{F}} \mathcal{E}\right\|_2$ and $\Omega^*\left(\frac1T\widetilde{W}^\top \mathcal{E}\right)$ in probability. This allows to obtain the following corollary, which states the rate of convergence of the prediction error of our estimator. 
	\begin{Corollary}\label{co.rate}
		Under Assumptions \ref{as.factors}, \ref{as.moments}, \ref{as.mixing} and \ref{as.rate}, and letting $\lambda= 2\Omega^*\left(\frac1T\widetilde{W}^\top \mathcal{E}\right)$, we have
		\begin{align*}\frac{1}{T}\left\|W\widehat{\delta}+\widehat{F}\widehat{\gamma}- W\delta-F\gamma\right\|_2^2&= O_P\left(s_\mu h_T^2+ \|A\|_2^2 +\frac{1}{p_x} \right).
		\end{align*}
	\end{Corollary}
	When $s_\mu h_T^2p_x\to \infty$, the term $\frac{1}{p_x}$ in the rate of convergence of Corollary \ref{co.rate} becomes negligible and our rate of convergence becomes the same as that of \cite{babii2022machine}. Hence, the condition $s_\mu h_T^2p_x\to \infty$ guarantees that the error in estimating the factors does not influence the rate of convergence of the prediction errors. This condition requires that $p_x$ grows sufficiently quickly with respect to $T$.  The standard rate of convergence of the LASSO estimator in prediction error would replace $h_T^2$ by $\log(p)/T$. We have $h_T^2$ in our rate because of the presence of $\tau$-mixing and polynomial tails.
	
	\section{Monte Carlo simulations}\label{sec.sim}
	
	In this section, we present a Monte Carlo study to examine the finite sample performance of our proposed method. We simulate data using the following data generating process. We use model \eqref{model_reg}-\eqref{model_m}, where $\rho_0 = 0.5$, $J = 2$, $\rho_1 = 0.2$, $\rho_2 = 0.1$ are fixed across all scenarios. We consider the sample sizes $T \in \{50, 100, 200\}$, which are representative of typical settings in nowcasting applications. We study several scenarios, with different degrees of cross-sectional dependence and fatness of the tails of the variables.
	
	The error term $\varepsilon_t$ follows a mean-zero AR(1) process with an autocorrelation coefficient of $0.4$ and variance $0.1$. The error in this AR(1) process follows either a Gaussian or a student-$t(5)$ distribution.
	
	The panel $\widetilde{z}$ represents high-frequency regressors that enter the model only through a sparse pattern. These variables are simulated at a weekly frequency assuming $m_z= 13$. We set $p_z = 30$ and the weekly variables are generated as AR(1) processes with an autocorrelation coefficient of $\rho_z=0.4$ and mean-zero error terms that have either a Gaussian or a student-$t(5)$ distribution. The error terms have a cross-sectional covariance matrix specified as $(1-\rho_z^2)\Sigma$, where $\Sigma^z_{ij} = c_z^{|i-j|}$ for $i, j \in [p_z]$, where $c_z\in\{0.1,0.4\}$ controls the degree of cross-sectional dependence among the regressors. We set $\omega_{z,1}$ and $\omega_{z,2}$ equal to the Beta$(1, 2)$ and Beta$(2, 2)$ densities, respectively. The other variables in $\widetilde{z}$ do not enter the model, that is $\omega_{z,k}\equiv 0$ for all $k\ge 3 $. 
	
	Next, the factors in $\widetilde{f}$ are monthly variables, that is $m_x=3$. We fix the number of factors to $K_f=1$ and simulate the factor as a zero-mean AR(1) process with an autoregressive coefficient of $0.4$ and a mean-zero error with unit variance which follows either a Gaussian or a student-$t(5)$ distribution.  Given these factors, we generate the $p_x=130$ monthly regressors in $\widetilde{x}$ according to the factor model \eqref{fac_model},
	where the loadings $\left\{\widetilde{b}_{k ,r},\ k\in[K_{x}],r \in[K_{f}]\right\}$ are i.i.d. uniform random variables on the interval $[-1,1]$. The idiosyncratic components $\left\{\widetilde{u}_{t-(j-1)/m_{x},k},\  t\in [T],k \in[K_{x}],j\in[m_{x}] \right\}$  are modeled as zero-mean autoregressive processes with an autoregressive coefficient of $\rho_u=0.4$, with error terms following a Gaussian or a student-$t(5)$ distribution with cross-sectional covariance matrix $\Sigma\left(1-\rho_u^2\right)$. The matrix $\Sigma$ is defined as $\Sigma_{ij} = c_u^{|i-j|}$ for $i, j \in [p_x]$, where $c_u\in\{0.1,0.4\}$ controls the degree of cross-sectional dependence among the monthly regressors. For $\omega_{x,1}$ and $\omega_{x,2}$ we use Beta$(1, 2)$ and Beta$(2, 2)$ densities, respectively. For all $k\ge 3 $,  $\omega_{x,k}\equiv 0$, that is only the first two variables in $\widetilde{x}$ enter the model.
	
	For the model estimation, we use Legendre polynomials of degree 3 to approximate the Beta density functions used in the weighting process ($D=4$). This choice ensures a flexible yet accurate representation of the underlying relationships. It is important to note that the model is simulated based on Beta density functions which we aim to approximate, hence the regressors that enter the sg-LASSO estimator are estimated based on Legendre polynomials. Lastly, the the number of factors is estimated using the growth ratio estimator of \cite{ahn2013eigenvalue}. The methods evaluated in our study correspond to those described in Section \ref{sec:methods}.

	Table \ref{OA-tab:mcs} in the Online Appendix  presents the mean squared error (MSE) results of three different methods—M1 (FAMIDAS), M2 (sg-LASSO-MIDAS), and M3 (sg-LASSO-FAMIDAS)—under various simulation settings. Panel A reports results for the baseline scenario with weak cross-sectional dependence ($c_z = c_u = 0.1$), while Panel B shows results for a scenario with stronger cross-sectional dependence ($c_z = c_u = 0.4$). In both panels, the performance is evaluated under two types of error distributions: Gaussian and student-$t(5)$. The reported MSE values are based on out-of-sample nowcasts.
	
	Across all sample sizes $T\in \{50, 100, 200\}$, M3 (sg-LASSO-FAMIDAS) consistently achieves the lowest MSE compared to M1 and M2, indicating superior predictive accuracy. As the sample size increases, all methods demonstrate improved performance (i.e., lower MSE), but the gains are most pronounced for M3. For example, in the baseline scenario (Panel A), M3’s MSE decreases from 1.1760 to 0.6133 under the Gaussian errors when $T$ increases from 50 to 200. In the stronger cross-sectional dependence scenario (Panel B), the relative performance of M3 remains robust, maintaining the lowest MSE across all settings. Notably, under the student-$t(5)$ distribution, M3 shows a significant relative improvement over M1 and M2, particularly for smaller sample sizes ($T = 50$), suggesting that the approach is more resilient to heavy-tailed errors when the true data generating process is sparse plus dense. 
	
	\tcr{In addition, we implemented designs where we set $\omega_{z,k} = \omega_{x,k} = 0$ (called the ``dense DGP'') and $\omega_{f,k} = 0$ (called the ``sparse DGP''). The results are reported in the Online Appendix, Tables \ref{OA-tab:mcs_dense} and \ref{OA-tab:mcs_sparse}, respectively. These results show that our proposed sparse plus dense approach outperforms sparse-only and dense-only approaches at almost no cost. Specifically, under the dense DGP, our method performs similarly to FAMIDAS, while under the sparse DGP, it achieves performance comparable to or better than sg-LASSO-MIDAS, which explicitly imposes sparsity without dense component. Note that the performance gains in the sparse DGP arise only when cross-sectional dependence is strong, likely because the factor component absorbs this dependence, thereby improving the estimation of the sparse signals—see, e.g., \citet{fan2020factor}.}
	
	Overall, the results indicate that sg-LASSO-FAMIDAS (M3) outperforms the other methods in terms of MSE, especially as the sample size increases or when errors exhibit heavy tails, highlighting its effectiveness in handling different data-generating scenarios.
	
	\section{Nowcasting GDP growth}\label{sec.now}
	
	For our empirical application, we focus on nowcasting current-quarter U.S. GDP growth, evaluating performance at three monthly horizons: two months before the end of the quarter ($h=2$), one month ahead ($h=1$), and at quarter-end ($h=0$). Since GDP data is typically released with a delay—often a month or more after the quarter ends—we mimick realistic nowcasting scenarios. For example, at the end of January (Q1), the model is trained on data until December (Q4), using available high-frequency data from January to nowcast Q1 GDP. The parameter $h$ indicates the number of months remaining in the target quarter. We use $m_z$ and $m_x$ to denote the number of recent high-frequency observations used from financial and macroeconomic panels, respectively. For monthly macro predictors, we include data from the current and previous quarters: $m_x = 6$ when $h=0$, $m_x = 5$ when $h=1$, and $m_x = 4$ when $h=2$. For weekly financial data, this translates to $m_z = 26$, $21$, and $17$ weeks, respectively, aligning with the number of weeks available from both the current and previous quarters.
	
	\subsection{Data}\label{subsec:data}
	
	We apply our methods to nowcast the initial, or ``advance," release of real U.S. GDP growth, which is typically published near the end of the first month of the following quarter. (An exception is the 2018 Q4 release, which was delayed due to a government shutdown.) Further details on this event are available from the Bureau of Economic Analysis. For our analysis, we use real-time GDP vintages from the ALFRED database maintained by the St. Louis FED. The in-sample period spans 1984 Q1 to 2007 Q4, with nowcasting beginning in 2008 Q1 and proceeding quarter by quarter using an expanding window approach.
	
	\medskip
	
	\noindent\textbf{Monthly macroeconomic regressors $\boldsymbol{\tilde{x}}$.} The monthly covariates and factors are drawn from the FRED-MD real-time dataset; see \citet{mccracken2016fred} for details. All variables are transformed according to the authors’ recommended procedures. We retain only those series that are consistently available across all vintages used for out-of-sample forecasting and exclude financial variables from the monthly panel. The final dataset includes 76 monthly macroeconomic series.
	
	\medskip
	
	\noindent\textbf{Weekly financial regressors $\boldsymbol{\tilde{z}}$.} We use weekly financial variables primarily drawn from those used to construct the Chicago Fed’s National Financial Conditions Index (NFCI). While the NFCI is a latent factor built from weekly, monthly, and quarterly data, we focus solely on the majority of its weekly components, excluding lower-frequency series. Our final panel includes 36 weekly financial series. Since financial data is available in real time, publication delays are not a concern. However, some series have shorter histories. To address this, we apply matrix completion with nuclear-norm regularization to impute missing values, ensuring a balanced panel. This imputation is performed at the original weekly frequency; see Section D of the Online Appendix for details.
	
	\medskip
	
	We provide the full list of series for monthly macro and weekly financial data with additional details in Section B of the Online Appendix.
	
	\subsection{Empirical results}\label{sec:res}
	
	Our candidate model set includes the AR(4) model, which we regard as the simplest and designate as our benchmark model. Consequently, our reported root mean squared forecast errors are presented relative to the AR(4). Notably, our conclusions remain consistent even when employing alternative benchmarks such as the random walk or AR(1). Our main results are based on the assumption that monthly macro covariates follow a factor model, hence entering the nowcasting equation with sparse plus dense signal. The weekly financial series are additional regressors that influence the target variables in a sparse way.  We have four autoregressive terms ($J=4$). As in the simulations, we use Legendre polynomials of degree 3 as the dictionary ($D=4$). In practice, we choose all tuning parameters by cross-validation adjusting for time series dependence, see, e.g., Section 4 in \cite{babii2022machine}. We estimate the number of factors using the growth ratio estimator of \cite{ahn2013eigenvalue}.
	
	Table \ref{tab:rmse} presents results for two sub-samples: Panel A covers the full out-of-sample period from 2008 Q1 to 2022 Q2, while Panel B focuses on the pre-COVID period ending in 2019 Q4. RMSEs for the autoregressive benchmark model are reported in absolute terms, with all other results expressed relative to this benchmark. As expected, the benchmark model exhibits a notable increase in prediction errors during the COVID-19 period, reflecting its inability to incorporate high-frequency information. Our results show that the factor-augmented sparse MIDAS regression substantially outperforms both FAMIDAS (dense) and sg-LASSO-MIDAS (sparse) methods over the full sample, with the largest gains observed during the pandemic. This underscores the effectiveness of combining sparse and dense components under heightened economic uncertainty. In contrast, prior to COVID-19, the performance of the sparse and factor-augmented sparse models is comparable, suggesting that the benefits of factor augmentation are most pronounced in volatile environments. Lastly, we note that the performance differences are statistically significant, as confirmed by the average superior predictive ability (aSPA) test of \citet{quaedvlieg2021multi}; see Section \ref{OA-sec:pvals} of the Online Appendix for details.
	
	\begin{table}[htp]
		\centering
		\begin{tabular}{rccc}
			& 2-month & 1-month & EoQ \\ 
			\hline
				&\multicolumn{3}{c}{Panel A. {\it Full sample}}\\
			AR(4) & 9.553 & 9.553 & 9.553  \\ 
			FAMIDAS & 0.590 & 0.495 & 0.580 \\ 
			sg-LASSO-MIDAS & 0.572 & 0.435 & 0.559\\
			sg-LASSO-FAMIDAS & 0.580 & 0.340 & 0.251  \\ 
			&\multicolumn{3}{c}{Panel B. {\it Up to COVID}}\\
			AR(4) & 1.934 & 1.934 & 1.934  \\ 
			FAMIDAS & 0.962 & 0.911 & 0.851 \\ 
			sg-LASSO-MIDAS & 0.827 & 0.837 & 0.733 \\ 
			sg-LASSO-FAMIDAS & 0.842 & 0.827 & 0.756 \\ 
			\hline\hline
		\end{tabular}
		\caption{Nowcast comparisons --- horizons are 2- and 1-month ahead, as well as the end of the quarter (EoQ). We report results for the full sample in Panel (A), and Panel (B) results excluding the COVID pandemic period, while Panel (C) reports results for the COVID pandemic period and beyond. The out-of-sample period starts from 2008 Q1 to 2022 Q2 (Panel A) and from 2008 Q1 to 2019 Q4 (Panel B). The RMSEs are reported relative to the AR model.\label{tab:rmse}}
	\end{table}
	
\begin{figure}[htp]
		\begin{subfigure}[b]{0.5\linewidth}
			\centering
			\includegraphics[width=1.0\columnwidth]{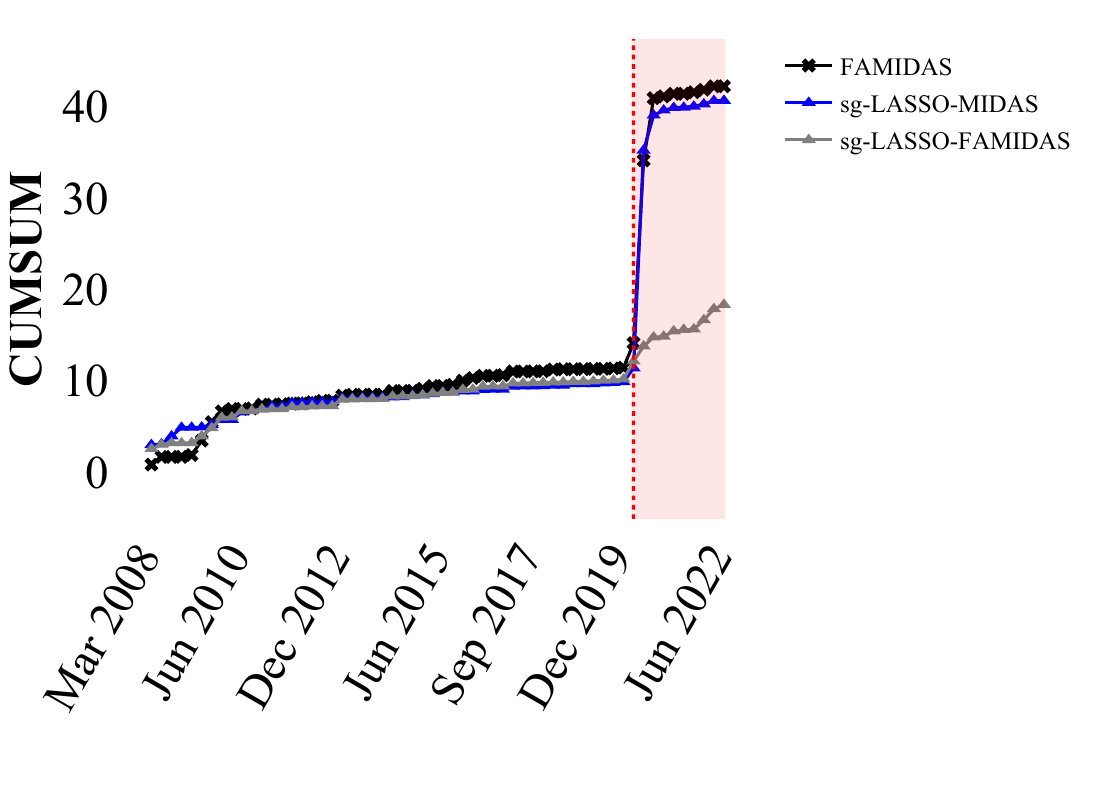}
			\caption{Full sample.}
			\label{fig:cumsum-precovid}
		\end{subfigure}
		\begin{subfigure}[b]{0.5\linewidth}
			\centering
			\includegraphics[width=1.0\columnwidth]{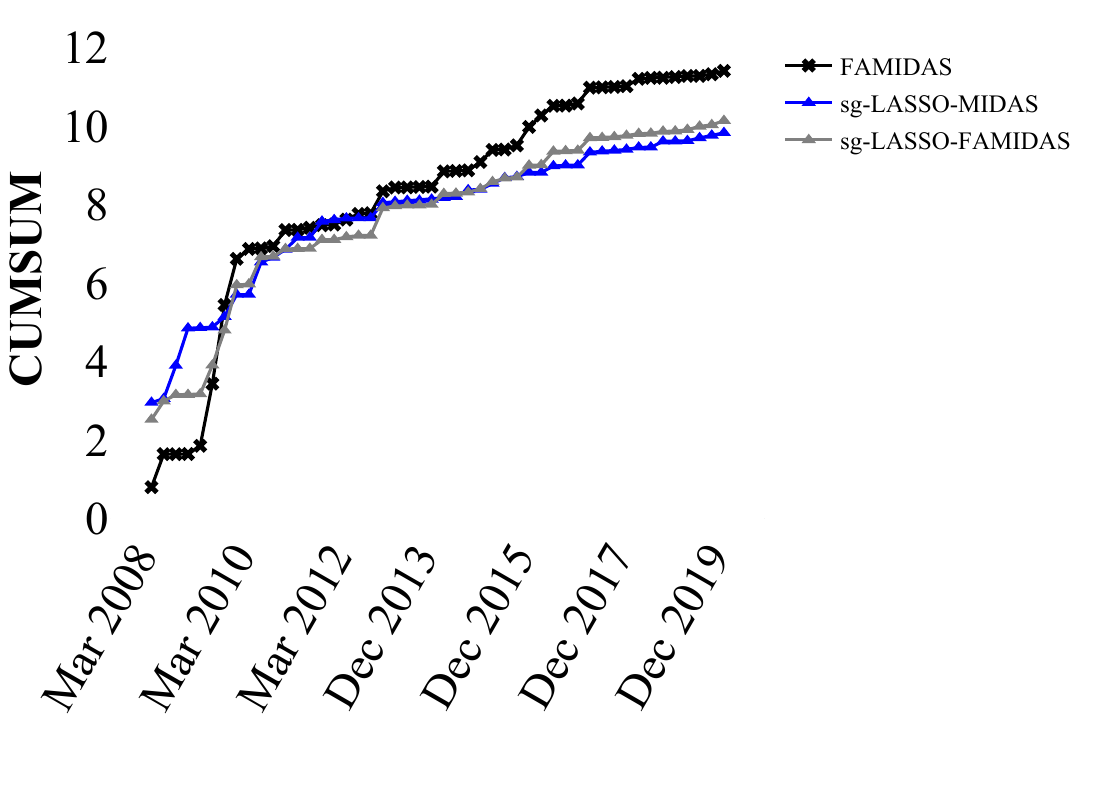}
			\caption{Up to COVID.}
			\label{fig:cumsum-covid}
		\end{subfigure}\hfill
		\caption{The figure illustrates the CUMSUM for the FAMIDAS, sg-LASSO-MIDAS and sg-LASSO-FAMIDAS using the whole out-of-sample period in Figure \ref{fig:cumsum-precovid}, while Figure \ref{fig:cumsum-covid} plots the pre-COVID pandemic period.  \label{fig:cumsum}}
	\end{figure}

	In Figure \ref{fig:cumsum}, we present the square-root cumulative sum of squared forecast errors (CUMSUM) for three competing methods across two sub-samples. These graphs are designed to visualize the performance differences between the models throughout the out-of-sample period. The forecast errors are calculated for the end-of-quarter horizon, and the CUMSUM is computed as follows:
	\begin{equation*}
		\text{CUMSUM}_{t,t+k} = \sqrt{\sum_{q=t}^{t+k} \hat \epsilon_{q,j}^2},
	\end{equation*}
	where $\hat \epsilon_{q,j}$, $j\in\{\text{FAMIDAS, sg-LASSO-MIDAS, sg-LASSO-FAMIDAS}\}$ are the out-of-sample nowcast errors. We plot the CUMSUM for the full sample period, corresponding to Panel A in Table \ref{tab:rmse}, and up to the COVID pandemic which corresponds to Panel B in the same Table.

	The plots highlight the differences between the methods, particularly during the onset of the COVID-19 outbreak. Prior to the pandemic, the performances of the sg-LASSO-FAMIDAS and sg-LASSO-MIDAS models are comparable, with the former providing more accurate nowcasts during the financial crisis and the latter performing better during stable periods between crises. This suggests that factor augmentation may not significantly impact performance during stable periods, despite the need to estimate additional unpenalized regression coefficients and the factors themselves. As the COVID-19 period unfolds, both the FAMIDAS and sg-LASSO-MIDAS methods show a marked decline in prediction accuracy, whereas the sg-LASSO-FAMIDAS method demonstrates notable resilience in handling the substantial shock introduced by the pandemic.
	
	Figure \ref{OA-fig:now} in the online appendix presents the nowcasts from the three forecasting methods across the three horizons, alongside the advance release of real GDP, which serves as the target. These plots, corresponding to the RMSEs in Table \ref{tab:rmse}, illustrate how each model tracks the advance estimate and how additional information improves accuracy over the quarter. We focus on the COVID-19 period, specifically Q2 and Q3 of 2020. In both quarters, the factor-augmented sparse regression model delivers highly accurate nowcasts by quarter-end. Since the advance estimate becomes available only a month into the following quarter, this highlights the effectiveness of timely data use. As expected, nowcast accuracy improves as more information becomes available. However, sg-LASSO-MIDAS, which uses only sparse signals, tends to understate the magnitude of GDP growth, while FAMIDAS, which relies solely on dense signals, produces more erratic forecasts. Overall, the figure shows that factor-augmented sparse MIDAS regression yields more balanced and reliable nowcasts during periods of high volatility.
	
	\subsection{\tcr{Economic rationale}}\label{sec:econ}
	
	\tcr{The results suggest that prior to the COVID pandemic, the data-generating process was predominantly sparse, whereas the COVID shock introduced a combination of idiosyncratic and common shocks in macroeconomic variables, but only idiosyncratic shocks in financial markets, which help nowcasting GDP (see the discussion and results backing the claim in a paragraph below). This pattern can be explained economically. The stringent physical restrictions during the pandemic had widespread effects on the real economy, generating common shocks across macro variables—see \citet{diebold2020covid}. In contrast, financial markets, although initially disrupted in early 2020, rebounded quickly, leading to a disconnect from real economic activity—a phenomenon widely discussed in policy circles (e.g., \citet{igan2020disconnect}). This disconnect and the absence of broad-based market co-movement help explain the lack of a dense financial component in our model. In the next two paragraphs, we discuss complementary analyses that confirm the aforementioned interpretation.}
	
	\tcr{First, we formally test for the presence of sparse and dense components. Using the data-driven bootstrap test of \citet{beyhum2024testing}, we test the null that the coefficients on the sparse component are zero. For the full real-time sample at the end-of-quarter horizon, the sparse part is significant (p = 0.002), and remains so pre-COVID (p = 0.037), indicating that idiosyncratic shocks contribute to explaining GDP. To test the significance of the macro factor (dense) component, we use the debiased LASSO with Gaussian multiplier bootstrap as in \citet{dezeure2017high}, where factors are not penalized in the initial LASSO regression. We find that PCA factors are significant in the full sample (p = 0.003) but not pre-COVID (p = 0.668). These test results mirror our out-of-sample findings: the sparse model performs best pre-COVID, while the combined sparse-plus-dense models improve substantially when including the COVID period. (We leave the theoretical validation of the tests by \citet{beyhum2024testing} and \citet{dezeure2017high} in our factor-augmented sparse regression setting with $\tau$-mixing data to further research.)}

	\tcr{We further assess sg-LASSO-FAMIDAS by replacing PCA-based macro factors with widely used observed indicators: the Aruoba-Diebold-Scotti index (ADS; \citealp{aruoba2009real}), the Chicago Fed National Activity Index (CFNAI; \citealp{brave2019new}), and the National Financial Conditions Index (NFCI; \citealp{brave2017introducing}, \citealp{amburgey2023real}). Using real-time vintages from the respective Federal Reserve Banks, we treat ADS and NFCI as weekly and CFNAI as monthly, preserving the original MIDAS structure and lag lengths. As shown in Table \ref{OA-tab:obsfactors} in the online appendix, ADS and CFNAI yield more accurate nowcasts than NFCI, with ADS performing best—likely due to its higher frequency and more stable inputs, consistent with \citet{diebold2020covid}. Since ADS and CFNAI track real economic activity, while NFCI captures financial conditions, these findings support our conclusion that macro variables carry both sparse and dense signals, whereas financial variables contribute primarily sparse signals. Notably, our PCA-based factor approach still outperforms these observed alternatives in predictive accuracy. Lastly, Table \ref{OA-tab:finspd} shows that adding PCA-based financial factors to the regression yields similar results, reinforcing our conclusion that financial variables mainly contribute sparse signals.}
	
	\subsection{\tcr{Robustness}}\label{sec:robust}
	
	\tcr{We conduct several robustness checks. First, we vary the factor model specification by imposing weak factors via sparse loadings following \citet{uematsu2022estimation} as well as the sparse PCA approach of \cite{zou2006sparse}, and assess alternative estimators for the number of factors, including those of \citet{bai2002determining} and \citet{bai2019rank}. We also evaluate a methodology where PCA factors are first computed from high-frequency macro predictors, then included in a MIDAS regression to nowcast GDP, as in, e.g.,  \citet{andreou2020mixed}. To assess sensitivity to structural breaks, we implement a rolling window scheme. }
	
	\tcr{Results appear in Tables \ref{OA-tab:pca_diff}, \ref{OA-tab:bn} and \ref{OA-tab:rolling} in the online appendix. Estimating factors from high-frequency data does not improve performance; the baseline sg-LASSO-FAMIDAS—where factors are extracted from MIDAS-weighted regressors—consistently outperforms the alternatives. For factor selection, the eigenvalue growth ratio of \citet{ahn2013eigenvalue} delivers the best results, with the eigenvalue ratio estimator yielding similar performance. Notably, regardless of the selection method, the sparse-plus-dense specification consistently outperforms sparse-only and dense-only models, suggesting it captures additional—though slightly noisier—predictive signals that improve the accuracy. Rolling window results (Table \ref{OA-tab:rolling}) align with expanding window results, indicating little impact of structural breaks, consistent with, e.g., \citet{boot2020does}.}
		
	\section{Conclusion}\label{sec.ccl}
	
	In this paper, we propose a factor-augmented sparse MIDAS regression for high-dimensional mixed-frequency data. By combining sparse and dense components, the method captures complex cross-frequency dependencies. We establish its theoretical properties under $\tau$-mixing and allow for model misspecification due to approximate sparsity or imperfect MIDAS lag structures, extending the literature with new convergence results for factor-augmented sparse regression.
	
	Monte Carlo and empirical results on nowcasting U.S. GDP growth show that the method outperforms purely sparse or dense models, especially during volatile periods such as the COVID-19 pandemic. Sparse signals dominate in stable times, while dense factor information is crucial under stress. The approach offers a robust framework for forecasting in complex environments and could be extended to nonlinear models or alternative high-frequency indicators.
	
	\if1\blind
	{	\vspace{-2.5em}
		{\section*{Funding sources}
			Jad Beyhum was financially supported by the Research Fund KU Leuven through the grant STG/23/014. Jonas Striaukas gratefully acknowledges the financial support from the European Commission, MSCA-2022-PF Individual Fellowship, Project 101103508, MACROML.
		} 
	} \fi
	\if0\blind
	{
	} \fi

	\vspace{-4.5em}
	
	\section*{Supplementary material}
	The Online Appendix contains the proofs, details on the data and other empirical results.
	
	\vspace{-2em}
		
	\bibliographystyle{agsm}
	\bibliography{paper}
	
	\clearpage

	\renewcommand{\theequation}{B.\arabic{equation}}
	\renewcommand\thetable{B.\arabic{table}}
	\renewcommand\thefigure{B.\arabic{figure}}
	\renewcommand\thesection{\Alph{section}}
	
\end{document}


\def\spacingset#1{\renewcommand{\baselinestretch}%
		{#1}\small\normalsize} \spacingset{1}

	
	\if1\blind
	{
		\title{ Online appendix of ``Factor-augmented sparse MIDAS regressions with an application to nowcasting''}
		\author{Jad Beyhum\\\vspace{0.5em}
			Department of Economics, KU Leuven, Belgium\vspace{0.5em}\\
			Jonas Striaukas
			\hspace{.2cm}\\
			Department of Finance, Copenhagen Business School, Denmark}
		\maketitle
	} \fi
	
	\if0\blind
	{
		\bigskip
		\bigskip
		\bigskip
		\begin{center}
			{\LARGE\bfPenalty }
		\end{center}
		\medskip
	} \fi

	\spacingset{1.3} 
	
	\renewcommand*{\thesection}{\Alph{section}}

	{\hypersetup{linkbordercolor=black,linkcolor = black,
			urlcolor  = blue,
			filecolor=black}
		\tableofcontents
	}
	\clearpage
	\setcounter{page}{1}
	\setcounter{section}{0}
	\setcounter{equation}{0}
	\setcounter{table}{0}
	\setcounter{figure}{0}
	\renewcommand{\theequation}{OA.\arabic{equation}}
	\renewcommand\thetable{OA.\arabic{table}}
	\renewcommand\thefigure{OA.\arabic{figure}}
	\renewcommand\thepage{OA. - \arabic{page}}
	\renewcommand\thealgorithm{OA.\arabic{algorithm}}
	
	\section{Proofs} 
	This section is organized as follows. First, Section \ref{sec.pfrate} contains the proof of Theorem \ref{P-th.rate}. Next, Corollary \ref{P-co.rate} is proved in Section \ref{sec.pcrate}. Then, the proof of Lemma \ref{P-lm.fac.est} can be found in Section \ref{sec.fac.est}. The proofs of Theorem \ref{P-th.rate}, Corollary  \ref{P-co.rate} and Lemma \ref{P-lm.fac.est} rely on some auxiliary lemmas presented and proved in Section \ref{sec.lm} and some pre-existing results restated in Section \ref{sec.pre}. 
	
	We will use the following notation. First, let $Y=(y_1,\dots,y_T)^\top$ and $\widehat{\Sigma}=\widetilde{W}^\top\widetilde{W}/T$. Next, for a matrix $H$, $\|H\|_{op}=\sigma_1(H)$ is its operator norm. For two matrices $H^{(1)},H^{(2)}$ of size $n_1\times n_2$, their scalar product is $\langle H^{(1)},H^{(2)}\rangle= \sum_{i=1}^{n_1} \sum_{j=1}^{n_2}H^{(1)}_{i,j}H^{(2)}_{i,j} $. For $N\in \mathbb{N}$, $I_N$ is the identity matrix of size $N\times N$.

	\subsection{Proof of Theorem \ref{P-th.rate}} \label{sec.pfrate}
	Throughout the proof, we work on the event 
	$$\left\{\left\|\widehat\Sigma - \Sigma\right\|_\infty \le \frac{\nu}{32s_\mu G^*}\right\},$$ which has probability going to $1$ by Lemma \ref{lm.T} \eqref{lTi}  and Assumption \ref{P-as.rate} \eqref{P-ratei}.
	First, remark that, by definition of $M_{\widehat{F}}$ and the Frisch-Waugh-Lovell theorem, we have 
	$$\widehat{\delta}\in\argmin_{d\in\R^{p}}\frac{1}{T}\left\|\widetilde{Y}-\widetilde{W}d\right\|^2_2+ 2\lambda \Omega(d),
	$$
	where $\widetilde{Y} = M_{\widehat{F}}Y $ .
	By Fermat's rule, $\hat{\delta}$ therefore satisfies
	\begin{equation*}
		\frac1T\widetilde{W}^\top(\widetilde{W}\widehat\delta - \widetilde{Y}) + \lambda z^* = 0, \text{ for some } z^*\in\partial\Omega(\widehat\delta) ,
	\end{equation*}
	where $\partial \Omega(\widehat\delta)$ is the sub-differential of $d\mapsto \Omega(d)$ at $\widehat\delta$. Taking the inner product with $\delta-\widehat\delta$, we get
	\begin{equation}\label{261024}
		\begin{aligned}
			\frac1T\left\langle \widetilde{W}^\top(\widetilde{Y}-\widetilde{W}\widehat\delta ),\delta - \widehat\delta\right\rangle&  = \lambda\left\langle z^*,\delta - \widehat\delta\right\rangle  \leq \lambda\left\{\Omega(\delta)- \Omega(\widehat\delta)\right\},
		\end{aligned}
	\end{equation}
	where the last inequality follows from the definition of the sub-differential. Notice that, 
	\begin{equation}\label{2610241}
		\begin{aligned}
			&\left\langle \widetilde{W}^\top(\widetilde{Y}-\widetilde{W}\widehat\delta ),\delta - \widehat\delta\right\rangle\\
			&=\left\langle \widetilde{W}^\top(\widetilde{Y}-\widetilde{W}\delta ),\delta- \widehat\delta\right\rangle + \left\langle \widetilde{W}^\top \widetilde{W}(\delta-\widehat\delta),\delta - \widehat\delta\right\rangle \\
			&=  \left\langle \widetilde{W}^\top  \mathcal{E},\delta - \widehat\delta\right\rangle+\left\langle \widetilde{W}^\top (F\gamma+A), \delta- \widehat\delta\right\rangle+\left\|\widetilde{W}(\widehat\delta-\delta)\right\|^2_{2},
		\end{aligned}
	\end{equation}
	where we used $Y=W\delta+F\gamma+A+\mathcal{E}$ in the last equality.
	By definition of the dual norm and the inequality of Cauchy-Schwarz, \eqref{261024} and \eqref{2610241} yield
	\begin{equation}\label{eq:two_point}
		\begin{aligned}
			&\frac{1}{T}\left\|\widetilde{W}(\widehat\delta-\delta)\right\|^2_{2} - \lambda\left\{\Omega(\delta)- \Omega(\widehat\delta)\right\}\\
			&\le\frac1T\left\langle \widetilde{W}^\top \left(\widetilde{Y}-\widetilde{W}\delta\right), \widehat\delta-\delta\right\rangle \\
			&\le\Omega^*\left(\frac1T\widetilde{W}^\top \mathcal{E}\right)\Omega\left(\widehat\delta-\delta\right)+\frac{1}{T}\left\|\widetilde{W}(\widehat\delta-\delta)\right\|_{2}\left(\|A\|_2+ \left\|M_{\widehat{F}}F\gamma\right\|_2\right).
		\end{aligned}
	\end{equation}
	Using that $\lambda\ge 2\Omega^*\left(\frac1T\widetilde{W}^\top  \mathcal{E}\right)$, we get 
	\begin{equation}\label{2807250}\begin{aligned}&\frac{1}{T}\left\|\widetilde{W}\left(\widehat\delta-\delta\right)\right\|^2_{2} -  \lambda\left\{\Omega(\delta)- \Omega(\widehat\delta)\right\}\\
			&\le \frac{\lambda}{2}\Omega\left(\widehat{\delta}-\delta\right)+\frac{1}{T}\left\|\widetilde{W}(\widehat\delta-\delta)\right\|_{2}\left(\|A\|_2+\left\|M_{\widehat{F}}F\gamma\right\|_2\right).\end{aligned}\end{equation}
	Now, for $d\in\R^p$, define
	\begin{align*}
		\Omega_0(d)&=\mu \|d_{\mathcal{S}_0}\|_1 +(1-\mu)\sum_{G\in\mathcal{G}_0}\|d_G\|_2;\\
		\Omega_1(d)&=\mu\|d_{\mathcal{S}_0^c}\|_1 +(1-\mu)\sum_{G\in\mathcal{G}_0^c}\|d_G\|_2.
	\end{align*}
	Let also $\Delta = \widehat\delta - \delta$. Next, remark that $\Omega(d)=\Omega_0(d)+\Omega_1(d)$, $\Omega_1(\delta)=0$ and $\Omega_1(\widehat\delta)=\Omega_1(\Delta)$. This leads to
	\begin{equation*}\begin{aligned}
			\notag \Omega(\delta)-\Omega(\widehat\delta)&=\Omega_0(\delta)+\Omega_1(\delta)-\Omega_0(\widehat\delta)-\Omega_1(\widehat\delta)\\
			\notag&= \Omega_0(\delta+\Delta-\Delta)-\Omega_0(\widehat\delta)-\Omega_1(\Delta)\\
			\notag&\le  \Omega_0(\widehat\delta)+\Omega_0(\Delta)-\Omega_0(\widehat\delta)-\Omega_1(\Delta)\quad \text{(By the triangle inequality)}\\
			&=\Omega_0(\Delta)-\Omega_1(\Delta).
		\end{aligned}
	\end{equation*}
	Combining this and \eqref{2807250}, we get
	\begin{equation}\label{bel30} \begin{aligned} \frac{1}{T}\left\|\widetilde{W}\Delta\right\|^2_{2}& \le \lambda\left\{\Omega_0(\Delta)-\Omega_1(\Delta)\right\}\\
			&\quad + \frac{\lambda}{2}\Omega(\Delta)+ \frac{1}{T}\left\|\widetilde{W}\Delta\right\|_{2}\left(\|A\|_2+ \left\|M_{\widehat{F}}F\gamma\right\|_2\right).
	\end{aligned}\end{equation}
	We consider two cases. 
	
	\bigskip 
	
	\noindent \textit{Case 1: $\|A\|_2+ \left\|M_{\widehat{F}}F\gamma\right\|_2 \le \frac{1}{2} \left\|\widetilde{W}\Delta\right\|_2 $.} 
	In this case, \eqref{bel30} implies
	\begin{equation}\label{bel31} \frac{1}{2T}\left\|\widetilde{W}\Delta\right\|^2_{2} \le \lambda\left\{\Omega_0(\Delta)-\Omega_1(\Delta)\right\}+ \frac{\lambda}{2}\Omega(\Delta).\end{equation}
	Since the left-hand-side of \eqref{bel31} is positive this shows that $\Delta$ belongs to the cone $\mathcal{C}=\{d\in\R^p:\ \Omega_1(d)\le 3\Omega_0(d)\}.$ This yields\begin{equation}\label{eq:Omega_Sigma_bound}
		\begin{aligned}
			\Omega(\Delta) & \leq 4\Omega_0(\Delta) \\
			& \le \mu \|\Delta_{\mathcal{S}_0}\|_1 +(1-\mu)\sum_{G\in\mathcal{G}_0}\|\Delta_G\|_2\\
			& \leq 4\left(\mu\sqrt{|S_0|}\|\Delta_{S_0}\|_2 + (1-\mu)\sqrt{|\mathcal{G}_0|}\sqrt{\sum_{G\in\mathcal{G}_0}\|\Delta_G\|_2^2}\right) \\
			& \leq 4\sqrt{s_\mu}\|\Delta\|_2\\
			& \leq 4\sqrt{s_\mu\Delta^\top\Sigma\Delta/\nu},	
		\end{aligned}
	\end{equation}
	where we used inequality of Cauchy-Schwarz, Assumption \ref{P-as.rate} \eqref{P-rateii}, and the definition of $\sqrt{s_\mu}$. Notice also that 
	\begin{equation}\label{2201}
		\begin{aligned}
			\Delta^\top \Sigma \Delta&=\Delta^\top \widehat{\Sigma} \Delta+\Delta^\top(\Sigma-\widehat{\Sigma})\Delta\\
			&\le \frac1T\left\|\widetilde{W}\Delta\right\|^2_{2}+\Omega(\Delta)\Omega^{*}\left((\widehat\Sigma - \Sigma)\Delta\right)\\
			&\le \frac1T\left\|\widetilde{W}\Delta\right\|^2_{2}+\Omega(\Delta)^2G^*\left\|\widehat{\Sigma}-\Sigma\right\|_\infty\\
			&\le \frac1T\left\|\widetilde{W}\Delta\right\|^2_{2}+\Omega(\Delta)^2\frac{\nu}{32s_\mu}\quad \text{\Bigg(Because  $\left\|\widehat\Sigma - \Sigma\right\|_\infty \le \frac{\nu}{32s_\mu G^*} $\Bigg),}\\
		\end{aligned}
	\end{equation}
	where the second inequality is a consequence of the definition of the dual norm, and the third inequality follows from the fact that, by Lemma \ref{lm.bab3}, we have
	\begin{align*}
		\Omega^*\left((\widehat\Sigma - \Sigma)\Delta\right)&=\mu \left\|(\widehat\Sigma - \Sigma)\Delta\right\|_\infty+(1-\mu)\max_{G\in\mathcal{G}}\left\|\left(\left(\widehat\Sigma - \Sigma\right)\Delta\right)_G\right\|_2\\
		&\le \mu \left\|(\widehat\Sigma - \Sigma)\Delta\right\|_\infty+(1-\mu)\sqrt{G^*} \left\|\left(\widehat\Sigma - \Sigma\right)\Delta\right\|_\infty\\
		&\le \mu \left\|\widehat\Sigma - \Sigma\right\|_\infty\|\Delta\|_1+(1-\mu)\sqrt{G^*}\left\|\widehat\Sigma - \Sigma\right\|_\infty\|\Delta\|_1\\
		&\le \mu \left\|\widehat\Sigma - \Sigma\right\|_\infty\|\Delta\|_1+(1-\mu)G^*\left\|\widehat\Sigma - \Sigma\right\|_\infty\|\Delta\|_{2,1}\\
		&\le G^*\left\|\widehat\Sigma - \Sigma\right\|_\infty\|\Delta\|_1,
	\end{align*}
	where in the first and third inequalities, we used the inequality of Cauchy-Schwarz and, in the second inequality, we used Hölder's inequality.
	Combining \eqref{eq:Omega_Sigma_bound} and \eqref{2201}, we obtain
	\begin{equation}\label{oa6}
		\begin{aligned}
			\Omega(\Delta)^2&\le \frac{16s_\mu}{\nu} \left(\frac1T\left\|\widetilde{W}\Delta\right\|^2_{2}+\Omega(\Delta)^2\frac{\nu}{32s_\mu}\right).
		\end{aligned}
	\end{equation}
	Using \eqref{bel31}, we get
	\begin{equation}\label{270824}
		\frac{1}{2T}\left\|\widetilde{W}\Delta\right\|^2_{2}  \le  \lambda\left\{\Omega_0(\Delta)-\Omega_1(\Delta)\right\}+ \frac{\lambda}{2}\Omega\left(\Delta\right)\le \frac32\lambda \Omega(\Delta),
	\end{equation}
	so that \eqref{oa6} implies 
	\begin{align*}
		\Omega(\Delta)^2&\le  \frac{16s_\mu}{\nu} \left(3\lambda\Omega(\Delta)+\Omega(\Delta)^2\frac{\nu}{32s_\mu}\right).
	\end{align*}
	This yields 
	\begin{equation}\label{omega15}\Omega(\Delta)\le \frac{96s_\mu\lambda}{\nu  }.\end{equation}
	Moreover, by \eqref{270824}, we obtain \begin{equation}\label{e15} \frac{1}{T}\left\|\widetilde{W}\Delta\right\|^2_{2}\le 3\lambda \Omega(\Delta)\le \frac{288s_\mu\lambda^2}{\nu  }.\end{equation}
	
	\bigskip 
	
	\noindent\textit{Case 2: $\|A\|_2+ \frac{1}{\sqrt{T}}\left\|M_{\widehat{F}}F\gamma\right\|_2> \frac{1}{2} \left\|\widetilde{W}\Delta\right\|_2 $.} This direcly yields 
	\begin{equation}\label{1310241}
		\frac{1}{T}\left\|\widetilde{W}\Delta\right\|^2_{2} \le \frac4T\left(\|A\|_2+ \left\|M_{\widehat{F}}F\gamma\right\|_2\right)^2.
	\end{equation}

	\bigskip

	\noindent\textit{Conclusion of the proof of Theorem \ref{P-th.rate}.} Combining \eqref{e15} and \eqref{1310241}, we get
	\begin{equation}\label{2708242}
		\begin{aligned}
			\frac{1}{T}\left\|\widetilde{W}\Delta\right\|^2_{2} &\le \frac{288s_\mu\lambda^2}{\nu  }+ \frac4T\left(\|A\|_2+\left\|M_{\widehat{F}}F\gamma\right\|_2\right)^2.
		\end{aligned}
	\end{equation}
	Next, remark that
	\begin{align}\label{81124} \widehat{F}\widehat{\gamma}&=P_{\widehat{F}} \left( Y-W\widehat{\delta}\right)=P_{\widehat{F}}W\left(\delta-\widehat{\delta}\right)+P_{\widehat{F}}F\gamma+ P_{\widehat{F}} \mathcal{E}.\end{align}
	Hence, we have 
	\begin{align*}
		W\widehat{\delta}+\widehat{F}\widehat{\gamma}- W\delta-F\gamma &= M_{\widehat{F}}\left(W\widehat{\delta}+\widehat{F}\widehat{\gamma}- W\delta-F\gamma\right)+ P_{\widehat{F}}\left(W\widehat{\delta}+\widehat{F}\widehat{\gamma}- W\delta-F\gamma\right)\\
		&=\widetilde{W}\left(\widehat{\delta}-\delta\right)-M_{\widehat{F}}F\gamma+P_{\widehat{F}}W\left(\widehat{\delta}-\delta\right) +\widehat{F}\widehat{\gamma}- P_{\widehat{F}}F\gamma\\
		&= \widetilde{W}\left(\widehat{\delta}-\delta\right)-M_{\widehat{F}}F\gamma+ P_{\widehat{F}} \mathcal{E},
	\end{align*}
	where we used \eqref{81124} in the last line.
	Therefore, we obtain
	\begin{align*}&\frac{1}{\sqrt{T}}\left\|W\widehat{\delta}+\widehat{F}\widehat{\gamma}- W\delta-F\gamma\right\|_2\\
		&\le \frac{1}{\sqrt{T}}\left(\left\|\widetilde{W}\left(\widehat{\delta}-\delta\right)\right\|_2+  \left\|M_{\widehat{F}}F\gamma\right\|_2+  \left\| P_{\widehat{F}} \mathcal{E}\right\|_2\right)\\
		&= \left(\frac{288s_\mu\lambda^2}{\nu  }+ \frac4T\left(\|A\|_2+\left\|M_{\widehat{F}}F\gamma\right\|_2\right)^2\right)^{1/2} +\frac{1}{\sqrt{T}}\left(  \left\|M_{\widehat{F}}F\gamma\right\|_2+  \left\| P_{\widehat{F}} \mathcal{E}\right\|_2\right).
	\end{align*}
	\subsection{Proof of Corollary \ref{P-co.rate}}\label{sec.pcrate}
	By Lemmas \ref{lm.T} \eqref{lTii} and \ref{lm.bab3}, we have
	$$ \lambda= 2\Omega^*\left(\frac1T\widetilde{W}^\top \mathcal{E}\right)=O_P\left(h_T+\frac{1}{p_x}\right).$$
	Plugging in this and the results of Lemma \ref{lm.T} \eqref{lTii}, \eqref{lTiii} and \eqref{lTiv} in the finite sample bound of Theorem \ref{P-th.rate}, we obtain 
	\begin{align*}\frac{1}{T}\left\|W\widehat{\delta}+\widehat{F}\widehat{\gamma}- W\delta-F\gamma\right\|_2^2&=
		O_P\left(s_\mu \left(h_T+\frac{1}{p_x}\right)^2+ \|A\|_2^2 +\frac{1}{T}+\frac{1}{p_x}\right)\\
		&=
		O_P\left(s_\mu h_T^2+s_\mu \frac{1}{p_x^2}+ \|A\|_2^2 +\frac{1}{T}+\frac{1}{p_x}\right)\\
		&= O_P\left(s_\mu h_T^2+ \|A\|_2^2 +\frac{1}{p_x}\right),
	\end{align*}
	where, in the second equality, we used Assumption \ref{P-as.rate} to simplify the rate.

	\subsection{Proof of Lemma \ref{P-lm.fac.est}}\label{sec.fac.est}
	
	\noindent\textit{Proof of \eqref{P-lfai}.}
	By Lemma \ref{lm.bai}, we have $\left\|\widehat{F}-FH^\top \right\|_2\le J_1+J_2+J_3,$ where 
	\begin{align*}
		J_1&= \frac1T \left\| FB^\top U^\top \widehat{F}V^{-1}\right\|_2;\\
		J_2&=\frac1T\left\| UBF^\top \widehat{F}V^{-1}\right\|_2;\\
		J_3&=\frac1T \left\| UU^\top  \widehat{F}V^{-1}\right\|_2.
	\end{align*}
	First, we bound $J_1$. It holds that 
	\begin{align*}
		J_1&\le \frac1T\|F\|_2 \|UB\|_2\left\|\widehat{F}\right\|_2\|V^{-1}\|_2\\
		&=O_P\left( \frac1T \sqrt{T} \sqrt{Tp_x} \sqrt{T} \frac{1}{p_x}\right)=O_P\left(\sqrt{\frac{T}{p_x}}\right),
	\end{align*}
	where we used Lemmas \ref{lm.F} \eqref{lfiii}, \eqref{lfvi} and \ref{lm.V} \eqref{lViii} and the fact that $\left\|\widehat{F}\right\|_2=\sqrt{RT}.$ Similarly, we have
	\begin{align*}
		J_2&\le \frac1T \|UB\|_2\|F\|_2\left \|\widehat{F}\right\|_2\|V^{-1}\|_2\\
		&=O_P\left( \frac1T \sqrt{Tp_x} \sqrt{T} \sqrt{T} \frac{1}{p_x}\right)=O_P\left(\sqrt{\frac{T}{p_x}}\right).
	\end{align*}
	Finally, it also holds that \begin{align*}J_3&\le \frac{1}{T}\left\| U U^\top\right\|_2\left\|  \widehat{F}\right\|_2 \| V^{-1}\|_2\\
		&=O_P\left( \frac1T\left(T\sqrt{p_x}+ p_x \sqrt{T}\right) \sqrt{T} \frac{1}{p_x}\right)=O_P\left(\sqrt{\frac{T}{p_x}}+1\right),\end{align*}
	where we used \ref{lm.F} \eqref{lfxvi} and \ref{lm.V} \eqref{lViii} and the fact that $\left\|\widehat{F}\right\|_2=\sqrt{RT}.$ Combining the bounds on $J_1$, $J_2$ and $J_3$, we obtain the result.\\
	
	\noindent\textit{Proof of \eqref{P-lfaii}.} By Lemma \ref{lm.bai}, we have $\left\|(\widehat{F}-FH^\top)^\top  \mathcal{E}\right\|_2 \le J_1+J_2+J_3,$
	where 
	\begin{align*}
		J_1&=  \frac1T \left\| \mathcal{E}^\top FB^\top U^\top \widehat{F}V^{-1}\right\|_2;\\
		J_2&=  \frac1T \left\| \mathcal{E}^\top UBF^\top \widehat{F}V^{-1}\right\|_2;\\
		J_3&=  \frac1T \left\| \mathcal{E}^\top UU^\top  \widehat{F}V^{-1}\right\|_2.
	\end{align*}
	We have 
	\begin{align*}
		J_1&\le \frac1T \left\| \mathcal{E}^\top F \right\|_2\left(\|UB\|_2 \left\|\widehat{F}-FH^\top\right\|_2+\|H\|_2\left\|B^\top U^\top F\right\|_2\right)\|V^{-1}\|_2\\
		&=O_P\left(\frac{1}{T} \sqrt{T}\left(\sqrt{Tp_x}\sqrt{\frac{T}{p_x} +1}+\sqrt{Tp_x}\right)\frac{1}{p_x}\right)=O_P\left( \frac{\sqrt{T}}{p_x}+\frac{1}{\sqrt{p_x}}\right),
	\end{align*}
	by Lemmas \ref{lm.F} \eqref{lfv}, \eqref{lfvi}, \eqref{lfix}, \ref{lm.H} \eqref{lHi} and \ref{lm.V} \eqref{lViii}, and statement \eqref{P-lfai}.
	Moreover, it holds that
	\begin{align*}
		J_2&\le \frac1T \| \mathcal{E}^\top UB\|_2\left\| F\right\|_2\left\|\widehat{F}\right\|_2\|V^{-1}\|_2\\
		&=O_P\left(\frac{1}{T} \sqrt{Tp_x}\sqrt{T}\sqrt{T}\frac{1}{p_x}\right)=O_P\left(\sqrt{\frac{T}{p_x}}\right),
	\end{align*}
	by Lemmas \ref{lm.F} \eqref{lfx}, \eqref{lfiii} and  \ref{lm.V} \eqref{lViii} and the fact that $\left\|\widehat{F}\right\|_2=\sqrt{RT}$.
	We also have 
	\begin{align*}
		\notag J_3&\le \frac1T \left\| \mathcal{E}^\top U\right\|_2\left(\|U\|_2 \left\|\widehat{F}-FH^\top\right\|_2+\left\|U^\top F\right\|_2\right)\|V^{-1}\|_2\\
		\notag &=O_P\left(\frac{1}{T}\sqrt{Tp_x}\left(\sqrt{Tp_x}\sqrt{\frac{T}{p_x}+1}+ \sqrt{Tp_x}\right)\frac{1}{p_x}\right)\\
		&=O_P\left(\sqrt{\frac{T}{p_x}}  +1\right),
	\end{align*}
	by Lemmas \ref{lm.F} \eqref{lfviii}, \eqref{lfii}, \eqref{lfvii}, \ref{lm.H} \eqref{lHi} and \ref{lm.V} \eqref{lViii}, and statement \eqref{P-lfai}.
	Combining the bounds on $J_1$, $J_2$ and $J_3$, we obtain \eqref{P-lfaii}.\\

	\noindent\textit{Proof of \eqref{P-lfaiii}.}
	By Lemma \ref{lm.bai}, we have $\left\|(\widehat{F}-FH^\top)^\top W\right\|_\infty \le J_1+J_2+J_3,$
	where 
	\begin{align*}
		J_1&=  \frac1T \left\|W^\top FB^\top U^\top \widehat{F}V^{-1}\right\|_\infty;\\
		J_2&=  \frac1T \left\|W^\top UBF^\top \widehat{F}V^{-1}\right\|_\infty;\\
		J_3&=  \frac1T \left\|W^\top UU^\top  \widehat{F}V^{-1}\right\|_\infty.
	\end{align*}
	We have 
	\begin{align*}
		\notag J_1&= \frac1T\max_{k\in[p], r\in[R]} \left|\sum_{j\in[R]}\left(W^\top F \right)_{k,j}\left(B^\top U^\top \widehat{F}V^{-1}\right)_{j,r}\right|\\
		&\le\frac{R}{T} \left\|W^\top F\right\|_\infty \left\|B^\top U^\top \widehat{F}V^{-1}\right\|_\infty \\
		&\le \frac{R}{T} \left\|W^\top F\right\|_\infty\left(\|UB\|_2 \left\|\widehat{F}-FH^\top\right\|_2+\|H\|_2\left\|B^\top U^\top F\right\|_2\right)\|V^{-1}\|_2\\
		&=O_P\left(\frac{1}{T} (Th_T+T)\left(\sqrt{Tp_x}\sqrt{\frac{T}{p_x} +1}+\sqrt{Tp_x}\right)\frac{1}{p_x}\right)\\
		&=O_P\left( \frac{T}{p_x}+\sqrt{\frac{T}{p_x}}\right),
	\end{align*}
	by Lemmas \ref{lm.F} \eqref{lfxi}, \eqref{lfvi}, \eqref{lfix}, \ref{lm.H} \eqref{lHi} and \ref{lm.V} \eqref{lViii}, and statement \eqref{P-lfai}.
	Moreover, it holds that
	\begin{align*}
		\notag J_2&=\frac1T\max_{k\in[p], r\in[R]} \left|\sum_{j\in[R]}\left(W^\top UB \right)_{k,j}\left(F^\top \widehat{F}V^{-1}\right)_{j,r}\right|\\
		&\le \frac{R}{T}  \left\|W^\top UB\right\|_\infty \left\|F^\top \widehat{F}V^{-1}\right\|_\infty \\
		&\le   \frac{R}{T}  \left\|W^\top UB\right\|_\infty  \left\| F\right\|_2\left\|\widehat{F}\right\|_2\|V^{-1}\|_2\\
		&=O_P\left(\frac{1}{T}\left(T\sqrt{p_x}h_T+T\right) \sqrt{T}\sqrt{T}\frac{1}{p_x}\right)\\
		&=O_P\left(\frac{T}{p_x}\left(\sqrt{p_x}h_T+1\right)\right),
	\end{align*}
	by Lemmas \ref{lm.F} \eqref{lfxiv}, \eqref{lfiii} and \ref{lm.V} \eqref{lViii}, and the fact that $\left\|\widehat{F}\right\|_2=\sqrt{RT}$.
	Finally, by the inequality of Cauchy-Schwarz, we also have 
	\begin{align*}
		\notag J_3&=\frac1T\max_{k\in[p],r\in[R]} \left|\sum_{\ell\in[p_x]}\left(W^\top U \right)_{k,\ell}\left(U^\top \widehat{F}V^{-1}\right)_{\ell,r}\right|\\
		&\le \frac1T \max_{k\in[p]} \left\|\sum_{t\in[T]}w_ {t,k} u_t \right\|_2\left\| U^\top \widehat{F}V^{-1}\right\|_2\\
		&\le \frac1T \max_{k\in[p]} \left\|\sum_{t\in[T]}w_ {t,k} u_t \right\|_2\left(\|U\|_2 \left\|\widehat{F}-FH^\top\right\|_2+\left\|U^\top F\right\|_2\right)\|V^{-1}\|_2\\
		\notag &=O_P\left(\frac{1}{T}\left(T\sqrt{p_x}h_T +T\right)\left(\sqrt{Tp_x} \sqrt{\frac{T}{p_x}+1} + \sqrt{Tp_x} \right)\frac{1}{p_x}\right)\\
		&=O_P\left(\left(\frac{T}{p_x}+\sqrt{\frac{T}{p_x}}\right) \left(\sqrt{p_x}h_T+1\right) \right),
	\end{align*}
	by Lemmas \ref{lm.F} \eqref{lfxv}, \eqref{lfii}, \eqref{lfvii} and \ref{lm.V} \eqref{lViii}, and statement \eqref{P-lfai}.
	Combining the bounds on $J_1$, $J_2$ and $J_3$, we obtain \eqref{P-lfaiii}.

	\subsection{Auxiliary lemmas}\label{sec.lm}

	\begin{Lemma}\label{lm.F} Under the assumptions of Lemma \ref{P-lm.fac.est}, the following holds:
		\begin{enumerate}[\textup{(}i\textup{)}]  
			\item\label{lfi}$ \left\|\frac1TF^\top F-I_R\right\|_{2}=O_P\left(\frac{1}{\sqrt{T}}\right)$;
			\item\label{lfii}$\left\|U\right\|_2=O_P\left(\sqrt{Tp_x}\right)$;
			\item\label{lfiii}$\|F\|_2=O_P(\sqrt{T})$;
			\item\label{lfiv}$\|  \mathcal{E}\|_2=O_P\left(\sqrt{T}\right)$;
			\item\label{lfv}$\left\|F^\top  \mathcal{E}\right\|_2=O_P\left(\sqrt{T}\right)$;
			\item\label{lfvi} $\|UB\|_2=O_P\left(\sqrt{Tp_x}\right)$;
			\item\label{lfvii}$\left\|F^\top U\right\|_2=O_P\left(\sqrt{Tp_x}\right)$;
			\item\label{lfviii}$\left\| \mathcal{E}^\top U\right\|_2=O_P\left(\sqrt{Tp_x}\right)$;
			\item\label{lfix} $\left\|F^\top UB\right\|_2=O_P\left(\sqrt{Tp_x}\right)$;
			\item\label{lfx} $\left\| \mathcal{E}^\top UB\right\|_2=O_P\left(\sqrt{Tp_x}\right)$;
			\item\label{lfxi} $\left\|W^\top F\right\|_\infty=O_P\left(Th_T +T\right);$
			\item\label{lfxii}  	$\left\|\frac1T W^\top W -\E[w_tw_t^\top]\right\|_\infty =O_P\left(\left(\frac{p^2}{T^{\kappa-1}}\right)^{1/\kappa}\vee \sqrt{\frac{\log(2p)}{T}}\right);$
			\item \label{lfxiii} $\left\|\frac1T W^\top F \left(\frac1T F^\top W \right) -\E[w_tf_t^\top] \E[w_tf_t^\top]^\top \right\|_\infty=O_P\left(h_T\right);$
			\item\label{lfxiv} $\left\|W^\top UB \right\|_\infty=O_P\left(T\sqrt{p_x}h_T+T\right);$
			
			\item\label{lfxv} $ \max_{k\in[p]} \left\|\sum_{t\in[T]}w_ {t,k} u_t \right\|_2=O_P\left(T\sqrt{p_x}h_T+T\right);$
			\item\label{lfxvi} $\left\|UU^\top\right\|_2=O_P\left(T\sqrt{p_x}+\sqrt{T}p_x\right);$
			\item\label{lfxvii} $\left\|W^\top  \mathcal{E}\right\|_\infty=O_P\left(h_T\right).$
		\end{enumerate}
	\end{Lemma}
	\begin{Proof} In this proof, we will repeatedly apply Lemma \ref{lm.bab1}. Its conditions hold by Assumptions \ref{P-as.moments} and \ref{P-as.mixing}.\\
		
		\noindent \textit{Proof of \eqref{lfi}.} This follows directly from Lemma \ref{lm.bab1} applied to $\zeta_t=f_tf_t^\top -\E[f_tf_t^\top]=f_tf_t^\top-I_R.$\\
		
		\noindent \textit{Proof of \eqref{lfii}.} We use that $\E[\|U\|_2^2]= \E[\sum_{t\in[T]} \sum_{k=1}^p u_{t,j}^2]=O(Tp)$ by Assumption \ref{P-as.moments}  \eqref{P-tailii} and Markov's inequality. \\
		
		\noindent \textit{Proof of \eqref{lfiii}.} We use that $\E[\|F\|_2^2]= \E[\sum_{t\in[T]} \sum_{r=1}^R f_{t,r}^2]=O(T)$ by Assumption \ref{P-as.moments}  \eqref{P-tailii} and Markov's inequality. \\ 
		
		\noindent \textit{Proof of \eqref{lfiv}.} We use that $\E[\| \mathcal{E}\|_2^2]= \E[\sum_{t\in[T]} \varepsilon_{t}^2]=O(T)$ by Assumption \ref{P-as.moments}  \eqref{P-tailii} and Markov's inequality. \\ 
		
		\noindent \textit{Proof of \eqref{lfv}.} Applying Lemma \ref{lm.bab1} to $\zeta_t=f_{t,r}\varepsilon_t$ gives $\sum_{t\in[T]}f_{t,r}\varepsilon_t=O_P(1/\sqrt{T}),$ which yields $\left\|F^\top  \mathcal{E}\right\|_2^2=\sum_{r=1}^R\left( \sum_{t\in[T]}  f_{t,r}\varepsilon_t\right)^2=O_P(T).$ \\ 
		
		\noindent\textit{Proof of \eqref{lfvi}.} We have 
		\begin{align*}
			\E[\|UB\|_2^2]&= p_x\sum_{t\in[T]} \sum_{r=1}^R \E\left[\left(p_x^{-1/2}\sum_{k=1}^{p_x}u_{t,k}b_{k,r}\right)^2\right]=O(Tp_x),
		\end{align*}
		by Assumption \ref{P-as.moments}  \eqref{P-tailii}.
		We obtain the result by Markov's inequality. \\

		\noindent\textit{Proof of \eqref{lfvii}.} We have 
		\begin{align*}
			\E\left[\left\|F^\top U\right\|_2^2\right]&=T \sum_{k=1}^{p_x}\sum_{r=1}^R \E\left[\left(T^{-1/2}\sum_{t=1}^{T}u_{t,k}f_{t,r}\right)^2\right]=O(Tp_x).
		\end{align*}
		We obtain the result by Markov's inequality. \\

		\noindent\textit{Proof of \eqref{lfviii}.} We have 
		\begin{align*}
			\E\left[\left\| \mathcal{E}^\top U\right\|_2^2\right]&=T \sum_{k=1}^{p_x} \E\left[\left(T^{-1/2}\sum_{t=1}^{T}u_{t,k}\varepsilon_t\right)^2\right]=O(Tp_x).
		\end{align*}
		We obtain the result by Markov's inequality. \\
		
		\noindent \textit{Proof of \eqref{lfix}.} This follows directly from Lemma \ref{lm.bab1} applied to $\zeta_t=f_t\left(p_x^{-1/2}\sum_{k=1}^{p_x}u_{t,k}b_{k,r}\right).$\\

		\noindent \textit{Proof of \eqref{lfx}.} This follows directly from Lemma \ref{lm.bab1} applied to $\zeta_t= \varepsilon_t\left(p_x^{-1/2}\sum_{k=1}^{p_x}u_{t,k}b_{k,r}\right).$\\
		
		\noindent \textit{Proof of \eqref{lfxi}.} By Lemma \ref{lm.bab1} applied to $\zeta_{t}=w_ {t} f_{t}^\top-\E[w_{t,k}f_{t,r}]$ and the triangle inequality, we have 
		\begin{align*}\frac1T\left\|W^\top F\right\|_{\infty} &= \max_{k\in[p],r\in[R]} \left|\frac{1}{T}\sum_{t\in[T]}w_ {t,k} f_{t,r} \right|\\
			&\le \max_{k\in[p],r\in[R]} \left|\frac{1}{T}\sum_{t\in[T]}w_ {t,k} f_{t,r}  -\E[w_{t,k}f_{t,r}]\right|+ \max_{k\in[p],r\in[R]} |\E[w_{t,k}f_{t,r}]|\\
			&=O_P\left(h_T\right)+ O(1).\end{align*}

		\noindent \textit{Proof of \eqref{lfxii}.}  This follows directly from Lemma \ref{lm.bab1} applied to $\zeta_t=w_tw_t^\top -\E[w_tw_t^\top]$.\\
		
		\noindent \textit{Proof of \eqref{lfxiii}.}  First, note that 
		\begin{equation}\label{2607242}\left\|\frac1T W^\top F\left(\frac1T F^\top W \right) -\E[w_tf_t^\top] \E[w_tf_t^\top]^\top \right\|_\infty\le J_1+J_2+J_3,
		\end{equation}
		where 
		\begin{align*}
			J_1&= 	\left\|\left(\frac1T W^\top F - \E[w_tf_t^\top] \right)  \E[w_tf_t^\top]^\top \right\|_\infty;\\
			J_2&=  \left\|\E[w_tf_t^\top] \left(\frac1T W^\top F - \E[w_tf_t^\top] \right)^\top    \right\|_\infty;\\
			J_3&= \left\|\left(\frac1T W^\top F - \E[w_tf_t^\top] \right)\left(\frac1T W^\top F - \E[w_tf_t^\top] \right)^\top    \right\|_\infty.
		\end{align*}
		Moreover, by Lemma \ref{lm.bab1} applied to $\zeta_t=w_tf_t^\top -\E[w_tf_t^\top]$, we have 
		\begin{equation}\label{2607241}
			\left\|\frac1T W^\top F - \E[w_tf_t^\top] \right\|_\infty =O_P\left(h_T\right).
		\end{equation}	
		This yields
		\begin{align*}
			J_1&=\max_{k,\ell\in[p_x]}\left|\sum_{r,j\in[R]}\left( \frac1T W^\top F - \E[w_tf_t^\top]\right)_{k,r}\left( \E[w_tf_t^\top]^\top\right)_{j,\ell}\right|\\
			&\le R \left\|\frac1T W^\top F - \E[w_tf_t^\top] \right\|_\infty \left\|\E[w_tf_t^\top] \right\|_\infty=O_P(h_T).
		\end{align*}
		Similarly, we have $J_2=O_P(h_T)$. Finally, by \eqref{2607241}, it also holds that 
		\begin{align*}
			J_3&=\max_{k,\ell\in[p_x]}\left|\sum_{r,j\in[R]}\left( \frac1T W^\top F - \E[w_tf_t^\top]\right)_{k,r}\left( \left(\frac1T W^\top F - \E[w_tf_t^\top]\right)^\top\right)_{j,\ell}\right|\\
			&\le R \left\|\frac1T W^\top F - \E[w_tf_t^\top] \right\|_\infty^2 =O_P(h_T^2)=o_P(h_T).
		\end{align*}
		Combining \eqref{2607242} and the bounds on $J_1,J_2$ and $J_3$.	\\

		\noindent \textit{Proof of \eqref{lfxiv}.} 
		By Lemma \ref{lm.bab1} applied to $$\zeta_t=w_ {t,k} \left(p_x^{-1/2}\sum_{k\in[p_x]} u_{t,k}b_k^\top \right) -\E\left[w_ {t,k} \left(p_x^{-1/2}\sum_{k\in[p_x]} u_{t,k}b_k^\top \right)\right],$$ we have 
		\begin{equation*}
			\max_{k\in[p],r\in[R]} \left|\frac1T \sum_{t\in[T]} w_ {t,k} \left(p_x^{-1/2}\sum_{\ell\in[p_x]} u_{t,\ell}b_{\ell,r} \right) -\E\left[w_ {t,k} \left(p_x^{-1/2}\sum_{\ell\in[p_x]} u_{t,\ell}b_{\ell,r} \right)\right] \right| =O_P\left(h_T\right).
		\end{equation*}	This implies that
		\begin{align*}\left\|W^\top UB \right\|_\infty&=T\max_{k\in[p],r\in[R]} \left|\frac1T \sum_{t\in[T]} w_ {t,k} \left(p_x^{-1/2}\sum_{\ell\in[p_x]} u_{t,\ell}b_{\ell,r} \right) \right|\\
			&\le T\sqrt{p_x} \max_{k\in[p]} \left\|\frac1T \sum_{t\in[T]} w_ {t,k} \left(p_x^{-1/2}\sum_{\ell\in[p_x]} u_{t,\ell}b_{\ell,r} \right) -\E\left[w_ {t,k} \left(p_x^{-1/2}\sum_{\ell\in[p_x]} u_{t,\ell}b_{\ell,r} \right)\right] \right\|_2\\
			& + T\max_{k\in[p]} \left\|\E\left[w_{t,k}\left(\sum_{k\in[p_x]} u_{t,k}b_k\right)\right]\right\|_2\\
			&=O_P\left(T\sqrt{p_x}h_T\right)+ O(T).\end{align*}
		
		\noindent \textit{Proof of \eqref{lfxv}.} 
		By Lemma \ref{lm.bab1} applied to $$\zeta_t=w_ {t} u_t^\top -\E\left[w_tu_t^\top\right],$$ we have 
		\begin{equation*}
			\left\|\frac1T \sum_{t\in[T]}w_ {t} u_t^\top -\E\left[w_tu_t^\top\right] \right\|_\infty=O_P\left(h_T\right).
		\end{equation*}
		We have 
		\begin{align*}&\max_{k\in[p]} \left\|\frac1T\sum_{t\in[T]}w_ {t,k}  u_t\right\|_2\\
			&\le \max_{k\in[p]} \left\|\frac 1T\sum_{t\in[T]}w_ {t,k} u_t -\E\left[w_{t,k}u_t\right]\right\|_2+ \max_{k\in[p]} \left\|\E\left[w_{t,k}u_t\right]\right\|_2\\
			&\le \sqrt{p_x} \left|\frac1T \sum_{t\in[T]} w_ {t} u_t^\top -\E\left[w_tu_t^\top\right] \right|_\infty+ \max_{k\in[p]} \left\|\E\left[w_{t,k}u_t\right]\right\|_2\\
			&=O_P\left(\sqrt{p_x}h_T\right)+ O(1).\end{align*}

		\noindent \textit{Proof of \eqref{lfxvi}.} It holds that 
		\begin{align*}
			\left\|UU^\top \right\|_2\le J_1+J_2,
		\end{align*}
		where
		\begin{align*}
			J_1&=\left\|UU^\top -\E\left[UU^\top\right] \right\|_2;\\
			J_2&= \left\|\E\left[UU^\top\right] \right\|_2.
		\end{align*}
		First, we bound $J_1$. 
		Since $\tcr{\max_{s,t\in[T]}\E\left[\left\{p_x^{-1/2}\left(u_s^\top u_t-\E[u_s^\top u_t]\right)\right\}^2\right]=O(1)}$, it holds that $$\E\left[\frac{1}{T^2}\sum_{s,t\in[T]} \left\{p_x^{-1/2}\left(u_s^\top u_t-\E[u_s^\top u_t]\right)\right\}^2\right]=O(1).$$
		By Markov's inequality, this yields
		\begin{align*}
			\left\|\frac{1}{T\sqrt{p_x}}\left(UU^\top -\E\left[UU^\top\right] \right)\right\|_2^2&=\frac{1}{T^2}\sum_{s,t\in[T]} \left\{p_x^{-1/2}\left(u_s^\top u_t-\E[u_s^\top u_t]\right)\right\}^2
			=O_P(1).
		\end{align*}
		This implies $J_1=O_P\left(T\sqrt{p_x}\right).$
		Now, we bound $J_2$.
		We have 
		\begin{equation}\label{260720252}
			\begin{aligned}
				\left\|\E\left[\frac{1}{\sqrt{T}p_x}UU^\top\right] \right\|_2^2&= \sum_{s,t\in[T]} \frac{1}{Tp_x^2}\E[u_s^\top u_t]^2\\
				&\le \max_{s,t\in[T]} \left|\frac{\E[u_s^\top u_t]}{p_x}\right| \max\limits_{s\in[T]}  \left(\sum_{t\in[T]}  \frac{1}{p_x}\left|\frac{\E[u_s^\top u_t]}{p_x}\right|\right).
			\end{aligned}
		\end{equation}
		We have 
		\begin{equation}\label{260720253}\max_{s,t\in[T]}\frac{1}{p_x}|\E[u_{s}^\top u_{t}]|\le \max_{s,t\in[T]}\max_{k\in[p_x]}\E[u_{s,k}u_ {t,j}]\le \max_{t\in[T]}\max_{k\in[p_x]}\E[u_ {t,j}^2]  =O(1) .\end{equation}
		Moreover, by Lemma \ref{lm.bab2}, for all $t\in[T],k\in[p]$, we have
		\begin{align*}\sum_{s\in[T]}  \frac{1}{p_x}|\E[u_{s,k} u_{t,k}]|\le \sum_{s\in\mathbb{N}}  \max_{k\in[p_x]}|\E[u_{0,k} u_{t,k}] |\le  \sum_{s\in\mathbb{N}} \tau_s^{\frac{q-2}{q-1}} \vertiii{u_{s,k}}_q^{\frac{q}{q-1}}=O(1),\end{align*}
		where the sum $\sum_{s\in\mathbb{N}} \tau_s^{\frac{q-2}{q-1}}\le \sum_{s\in\mathbb{N}} cs^{-a\frac{q-2}{q-1}}  $ converges since $a>(q-1)/(q-2)$ by Assumption \ref{P-as.mixing}. This, \eqref{260720252} and \eqref{260720253} imply that $J_2=O(\sqrt{T}p_x)$. Combining the bounds on $J_1$ and $J_2$, we obtain the result.\\
		
		\noindent \textit{Proof of \eqref{lfxvii}.} This follows directly from Lemma \ref{lm.bab1} applied to $\zeta_t= \varepsilon_tw_t.$
	\end{Proof}
	\begin{Lemma}\label{lm.V} Under the assumptions of Theorem \ref{P-th.rate}, 
		the following holds
		\begin{enumerate}[\textup{(}i\textup{)}]  
			\item\label{lVi} $\|U\|_{op}=O_P\left(\sqrt{T}p_x^{1/4}+\sqrt{p_x}T^{1/4}\right)$;
			\item\label{lVii} There exists $c>0$ independent of $T$ such that $$\P\left(\sigma_R\left(\frac{1}{Tp_x}BF^\top FB^\top\right)\ge c\right)\to 1;$$
			\item\label{lViii} With probability going to $1$, $V$ is invertible. Moreover, we have $\|V^{-1}\|_2=O_P\left(p_x^{-1}\right)$.
		\end{enumerate}
	\end{Lemma}
	\begin{Proof}
		\noindent\textit{Proof of \eqref{lVi}.} We have $\|U\|_{op}\le \sqrt{\|UU^\top\|_2}=O_P\left(\sqrt{T}p_x^{\frac14}+\sqrt{p_x}T^{\frac14}\right)$ by Lemma \ref{lm.F} \eqref{lfxvi}.\\

		\noindent\textit{Proof of \eqref{lVii}.} By Weyl's inequality, we have 
		\begin{equation}\label{2707250}\begin{aligned}&\left|\sigma_R\left(\frac{1}{Tp_x}BF^\top FB^\top\right)-\sigma_R\left(\frac{1}{p_x} BB^\top\right)\right|\\
				&\le \left\|\frac{1}{p_x}B\left(\frac{1}{T}F^\top F-I_R\right)B^\top \right\|_{op}\\
				&\le  \frac{1}{p_x}\|B\|_{op}^2 \left\|\frac1TF^\top F-I_R\right\|_{2},\\
				& \le   \frac{1}{p_x}\|B^\top B\|_{2}^2\left\|\frac1TF^\top F-I_R\right\|_{2}
				=O_P\left(\frac{1}{\sqrt{T}}\right),
			\end{aligned}
		\end{equation}
		where we used Lemma \ref{lm.F} \eqref{lfi} and Assumption \ref{P-as.factors} \eqref{P-f4iv}.
		Moreover, let $2c$ be the lower bound mentioned in Assumption \ref{P-as.factors} \eqref{P-f4ii}, then we have $\sigma_R\left(\frac{1}{p_x} BB^\top\right)\ge 2c$, so that $$\P\left(\sigma_R\left(\frac{1}{Tp_x}BF^\top FB^\top\right)\ge c\right)\to 1,$$
		by \eqref{2707250}.\\
		
		\noindent\textit{Proof of \eqref{lViii}.} First, by Weyl's inequality, we have 
		\begin{align*} &\left|\sigma_R\left(\frac{1}{T}X^\top X \right)-\sigma_R\left(\frac{1}{T}BF^\top FB^\top\right)\right|\\
			&\le  \frac{1}{T}\left\|X^\top X-BF^\top FB^\top\right\|_{op}\\
			&\le  \frac{1}{T}\left\|(FB^\top +U)^\top(FB^\top +U) -BF^\top FB^\top\right\|_{op}\\
			&\le \frac{1}{T} \left\|FB^\top U^\top + U BF^\top+U^\top U\right\|_{op}\\
			& \le  \frac2T \left\|FB^\top\right\|_{op} \left\| U \right\|_{op}+ \frac1T \left\|U^\top U\right\|_{op}\\
			&  \le  \frac2T \left\|F\right\|_{2} \|B\|_2\left\| U \right\|_{op}+ \frac1T \left\|U\right\|_{op}^2\\
			&= O_P\left(  \frac1T \sqrt{T}\sqrt{p_x} \left(\sqrt{T}p_x^{1/4}+\sqrt{p_x}T^{1/4}\right)+\frac1T\left(T\sqrt{p_x} +\sqrt{T}p_x\right)\right)\\
			&= o_P\left(p_x\right),
		\end{align*}
		where we used \eqref{lVi}, Lemma \ref{lm.F} \eqref{lfiii} and Assumption \ref{P-as.factors} \eqref{P-f4iv}.
		This shows that $\P\left(\sigma_R\left(\frac{1}{T}XX^\top\right)\ge cp_x/2\right)\to 1$, which also yields $\|V^{-1}\|_2\le \sqrt{R}\left(\sigma_R\left(\frac{1}{T}XX^\top\right)\right)^{-1}=O_P\left(p_x^{-1}\right)$.
	\end{Proof}
	\begin{Lemma}\label{lm.bai}Under the assumptions of Lemma \ref{P-lm.fac.est}, it holds that $$\widehat{F}-FH^\top = \frac1T FB^\top U^\top \widehat{F}V^{-1}+\frac1T  UBF^\top \widehat{F}V^{-1}+\frac1T UU^\top  \widehat{F}V^{-1}.$$
	\end{Lemma}
	\begin{Proof}
		Recall that $H=\frac1TV^{-1}\widehat{F}^\top FB^\top B$ and $\widehat{F}V=\frac1T XX^\top\widehat{F}$. As a result, we have
		\begin{align*}
			\widehat{F}V& = T^{-1}XX^\top\widehat{F}\\
			&=  \frac1T (FB^\top  + U)(FB^\top  + U)^\top\widehat{F}\\
			&= \frac1T FB^\top B F^\top\widehat{F}+\frac1T  FB^\top U^\top \widehat{F}+ \frac1T UBF^\top \widehat{F}+T^{-1}UU^\top  \widehat{F}.
		\end{align*}
		Multiplying both sides by $V^{-1}$, we get the result.
	\end{Proof}
	
	\begin{Lemma}\label{lm.H}Under the assumptions of Lemma \ref{P-lm.fac.est}, the following holds:
		\begin{enumerate}[\textup{(}i\textup{)}]  
			\item\label{lHi} $\|H\|_2=O_P(1)$;
			\item\label{lHii} $\|H^\top H-I_R\|_2=O_P\left(\frac{1}{\sqrt{T}}+\frac{1}{\sqrt{p_x}}\right)$.
		\end{enumerate}
	\end{Lemma}
	\begin{Proof} \noindent\textit{Proof of \eqref{lHi}.} Recall that $H=\frac1TV^{-1}\widehat{F}^\top FB^\top B$. This yields
		\begin{align*}
			\|H\|_2&\le \frac1T\|V^{-1}\|_2  \left\|\widehat{F}\right\|_2\|F\|_2\|B^\top B\|_2=O_P(1),
		\end{align*}
		by Lemmas \ref{lm.V} \eqref{lViii} and \ref{lm.F} \eqref{lfiii}, $\|\widehat{F}\|_2=\sqrt{T}$ and the fact that $\|B^\top B\|_2=O(p_x)$ by Assumption \ref{P-as.factors} \eqref{P-f4ii}.\\
		
		\noindent\textit{Proof of \eqref{lHii}.}
		We have 
		\begin{equation} \label{2707251}
			\begin{aligned} \left\|H^\top H-I_R\right\|_2&\le  \left\|H^\top H-H^\top \frac{F^\top F}{T}H \right\|_2+ \left\|H^\top \frac{F^\top F}{T}H -I_R\right\|_2\\
				&\le  \left\|H^\top \left(I_R-  \frac{F^\top F}{T}\right)H \right\|_2+ \left\|H^\top \frac{F^\top F}{T}H -I_R\right\|_2\\
				&\le \|H\|_2^2 \left\|\frac{F^\top F}{T}-I_R\right\|_2+ \left\|H^\top \frac{F^\top F}{T}H -I_R\right\|_2\\
				&=O_P\left(\frac{1}{\sqrt{T}}\right) + \left\|H^\top \frac{F^\top F}{T}H -I_R\right\|_2,
			\end{aligned}
		\end{equation}
		by Lemma \ref{lm.F} \eqref{lfi}.
		Note that 
		\begin{align*}H^\top \frac{F^\top F}{T}H-I_R&=H^\top \frac{F^\top F}{T}H-\frac{ \widehat{F}^\top\widehat{F}}{T}  \\
			&= \frac{1}{T}  \left(H F^\top -\widehat{F}^\top\right)FH +\frac1T \widehat{F}^\top \left(FH^\top-\widehat{F}\right).
		\end{align*}
		Hence, we have 
		\begin{align*}
			\left\|H^\top \frac{F^\top F}{T}H -I_R\right\|_2&\le \frac1T\left\|\widehat{F}-FH^\top\right\|_2\|F\|_2\|H\|_2+ \left\|\widehat{F}-FH^\top\right\|_2\|\widehat{F}\|_2\\
			&=O_P\left( \frac{1}{\sqrt{T}} +\frac{1}{\sqrt{p_x}} \right),
		\end{align*}
		where we used Lemmas \ref{P-lm.fac.est} \eqref{P-lfai} and \ref{lm.F} \eqref{lfiii} and $\|\widehat{F}\|_2=\sqrt{RT}$. (Note that \eqref{lHii} is not used to prove Lemma \ref{P-lm.fac.est} \eqref{P-lfai} so that we can indeed use the latter.) This and \eqref{2707251} show \eqref{lHii}.
	\end{Proof}

	\begin{Lemma}\label{lm.T}Under the assumptions of Theorem \ref{P-th.rate}, the following holds: 
		\begin{enumerate}[\textup{(}i\textup{)}]  
			\item\label{lTi} $\left\|\widehat{\Sigma}-\Sigma\right\|_\infty=O_P\left(\left(\left(\frac{p^2}{T^{\kappa-1}}\right)^{1/\kappa}\vee \sqrt{\frac{\log(2p)}{T}}\right)+\frac{1}{\sqrt{p_x}}\right)$;
			\item\label{lTii} $\frac{1}{T}\left\|\widetilde{W}^\top  \mathcal{E}\right\|_{\infty}=O_P\left(h_T+\frac{1}{p_x}\right)$;
			\item \label{lTiii}$\left\|M_{\widehat{F}}F\gamma\right\|_2=O_P\left(\sqrt{\frac{T}{p_x}+1}\right)$;
			\item \label{lTiv} $\left\|P_{\widehat{F}} \mathcal{E}\right\|_2=O_P\left(\sqrt{\frac{T}{p_x}+1}\right)$.
		\end{enumerate}
		
	\end{Lemma}
	\begin{Proof}
		\noindent \textit{Proof of \eqref{lTi}.} 
		We have 
		\begin{align*}
			\left\|\widehat{\Sigma}-\Sigma\right\|_\infty&\le\left\|\frac1T\widetilde{W}^\top \widetilde{W}-\Sigma\right\|_\infty\\
			&\le\left\|\frac1T W^\top\left(I_T-\frac1T\widehat{F}\widehat{F}^\top\right) W-\Sigma\right\|_\infty\\
			&\le J_1+J_2+J_3+J_4+J_5,
		\end{align*}
		where 
		\begin{align*}
			J_1&= \left\|\frac1T W^\top\frac1T(\widehat{F}-FH^\top)(\widehat{F}-FH^\top)^\top W\right\|_{\infty};\\
			J_2&= \left\|\frac1T W^\top\frac1T(\widehat{F}-FH^\top ) H F^\top W\right\|_{\infty};\\
			J_3&= \left\| \frac1T W^\top\frac1TFH^\top(\widehat{F}-FH^\top)^\top W\right\|_{\infty};\\
			J_4&= \left\|\frac1T W^\top \frac1T F \left(I_R-H^\top H \right)F^\top W\right\|_{\infty};\\
			J_5&= \left\|\frac1T W^\top \left(I_T-\frac1T F F^\top\right)W-\Sigma\right\|_{\infty}.
		\end{align*}
		Let us first bound $J_1$. It holds that 
		\begin{align*}
			J_1&=\frac{1}{T^2}\max_{k,\ell \in[p_x]}\left| \sum_{r\in[R]} \left(W^\top \left(\widehat{F}-FH^\top\right)\right)_{k,r}\left(\left(\widehat{F}-FH^\top\right)^\top W\right)_{r,\ell}\right| \\
			&\le \frac{R}{T^2}\left\|\left(\widehat{F}-FH^\top\right)^\top W\right\|_\infty^2\\
			&= O_P\left(\frac{1}{T^2}\left(\frac{T}{p_x}+\sqrt{\frac{T}{p_x}}\right)^2\left(\sqrt{p_x}h_T+1\right)^2\right)\\
			&= o_P\left(\frac{1}{p_x}+\frac{1}{T}\right),
		\end{align*}
		by Lemma \ref{P-lm.fac.est} \eqref{P-lfaiii}.
		Next, we consider $J_2$. We have 
		\begin{align*}J_2 & =\frac{1}{T^2}\max_{k,\ell \in[p_x]}\left| \sum_{r\in[R]} \left(W^\top \left(\widehat{F}-FH^\top\right)\right)_{k,r}\left(H F^\top W\right)_{r,\ell}\right|   \\
			&\le \frac{R}{T^2}\left\|\left(\widehat{F}-FH^\top\right)^\top W\right\|_\infty \left\|H F^\top W\right\|_\infty\\
			&\le \frac{R}{T^2}\left\|\left(\widehat{F}-FH^\top\right)^\top W\right\|_\infty\max_{r\in[R],k\in[p_x]}\left|\sum_{j\in[R] } H_{r,j} \left(F^\top W\right)_{j,k}\right|\\
			&\le \frac{R^2}{T^2}\left\|\left(\widehat{F}-FH^\top\right)^\top W\right\|_\infty\left\|H \right\|_2\|F^\top W\|_\infty\\
			&=O_P\left(\frac{1}{T^2}\left(\frac{T}{p_x}+\sqrt{\frac{T}{p_x}}\right)\left(\sqrt{p_x}h_T+1\right)\left(Th_T+T\right)\right)\\
			&=O_P\left(\frac{1}{\sqrt{p_x}}+\frac{1}{\sqrt{T}}\right),
		\end{align*}
		by Lemmas \ref{P-lm.fac.est} \eqref{P-lfaiii}, \ref{lm.H} \eqref{lHi} and \ref{lm.F} \eqref{lfxi}.
		By similar reasoning, it holds that $$ J_3=o_P\left(\frac{1}{\sqrt{p_x}}+\frac{1}{\sqrt{T}}\right).$$ Then, we focus on $J_4$. It holds that 
		\begin{align*}
			J_4&= \frac{1}{T^2}\max_{k,\ell \in[p_x]}\left| \sum_{r\in[R]} \left(W^\top F\right)_{k,r}\left(\left(I_R-H^\top H \right)F^\top W\right)_{r,\ell}\right|\\
			&\le \frac{R}{T^2}  \|W^\top F\|_\infty \|\left(I_R-H^\top H \right)F^\top W\|_\infty\\
			&= \frac{R}{T^2}  \|W^\top F\|_\infty \max_{r\in[R],k\in[p_x]}\left|\sum_{j\in[R] } \left(I_R-H^\top H \right)_{r,j} \left(F^\top W\right)_{j,k}\right|\\
			& \le \frac{R^2}{T^2} \|W^\top F\|_\infty^2  \left\| I_R-H^\top H\right\|_2\\
			&=O_P\left(\frac{1}{T^2} (Th_T+T)^2 \left(\frac{1}{\sqrt{T}}+\frac{1}{\sqrt{p_x}}\right)\right)\\
			&= O_P\left( \frac{1}{\sqrt{T}}+\frac{1}{\sqrt{p_x}}\right),
		\end{align*}
		by Lemmas \ref{lm.H} \eqref{lHii} and \ref{lm.F} \eqref{lfxi}.
		We finish by bounding $J_5$. We have 
		\begin{align*}
			J_5 &\le \left\|\frac1T W^\top W -\E[w_tw_t^\top]\right\|_\infty +\left\|\frac1T W^\top F \left(\frac1T F^\top W \right) -\E[w_tf_t^\top] \E[w_tf_t^\top]^\top \right\|_\infty\\
			&=O_P\left(\left(\frac{p^2}{T^{\kappa-1}}\right)^{1/\kappa}\vee \sqrt{\frac{\log(2p)}{T}}\right),
		\end{align*}
		by Lemmas \ref{lm.F} \eqref{lfxii} and \eqref{lfxiii}. Combining the bounds on $J_1$ to $J_5$, we obtain \eqref{lTi}.\\
		
		\noindent \textit{Proof of \eqref{lTii}.} We have 
		\begin{align*}
			\frac1T \left\|\widetilde{W}^\top  \mathcal{E}\right\|_{\infty}&= \frac1T\left\|T^{-1}W^\top \left(I_T-\frac1T\widehat{F}\widehat{F}^\top\right)   \mathcal{E}\right\|_{\infty}\\
			& \le J_1+J_2+J_3+J_4+J_5+J_6,
		\end{align*}
		where 
		\begin{align*}
			J_1&= \left\|\frac1T W^\top\frac1T(\widehat{F}-FH^\top)(\widehat{F}-FH^\top)^\top  \mathcal{E}\right\|_{\infty};\\
			J_2&= \left\|\frac1T W^\top\frac1T(\widehat{F}-FH^\top ) H F^\top  \mathcal{E}\right\|_{\infty};\\
			J_3&= \left\| \frac1T W^\top\frac1TFH^\top(\widehat{F}-FH^\top)^\top  \mathcal{E}\right\|_{\infty};\\
			J_4&= \left\|\frac1T W^\top \frac1T F \left(I_R-H^\top H \right)F^\top  \mathcal{E}\right\|_{\infty};\\
			J_5&= \left\|\frac{1}{T^2} W^\top FF^\top    \mathcal{E}\right\|_{\infty} ;\\
			J_6&=\left\|\frac1T W^\top   \mathcal{E}\right\|_{\infty}.\\
		\end{align*}
		Let us first bound $J_1$. It holds that 
		\begin{align*}
			J_1&=\frac{1}{T^2}\max_{k \in[p_x]}\left| \sum_{r\in[R]} \left(W^\top \left(\widehat{F}-FH^\top\right)\right)_{k,r}\left(\left(\widehat{F}-FH^\top\right)^\top  \mathcal{E}\right)_{r}\right|\\
			&\le \frac{R}{T^2}\left\|\left(\widehat{F}-FH^\top\right)^\top W\right\|_\infty\left\|\left(\widehat{F}-FH^\top\right)^\top  \mathcal{E}\right\|_\infty\\
			&= O_P\left(\frac{1}{T^2}\left(\frac{T}{p_x}+\sqrt{\frac{T}{p_x}}\right)\left(\sqrt{p_x}h_T+1\right)\sqrt{\frac{T}{p_x}+1}\right)\\
			&= o_P\left(\frac{1}{T\sqrt{p_x}} +\frac{1}{T^{3/2}}+\frac{1}{p_x\sqrt{T}}\right),
		\end{align*}
		by Lemma \ref{P-lm.fac.est} \eqref{P-lfaii} and \eqref{P-lfaiii}.
		Next, we consider $J_2$. We have 
		\begin{align*}J_2 & =\frac{1}{T^2}\max_{k\in[p_x]}\left| \sum_{r\in[R]} \left(W^\top \left(\widehat{F}-FH^\top\right)\right)_{k,r}\left(H F^\top  \mathcal{E}\right)_{r}\right| \\
			&\le \frac{R}{T^2}\left\|\left(\widehat{F}-FH^\top\right)^\top W\right\|_\infty \left\|H F^\top  \mathcal{E}\right\|_\infty\\
			&\le \frac{R}{T^2}\left\|\left(\widehat{F}-FH^\top\right)^\top W\right\|_\infty\left\|H \right\|_2\left\| F^\top  \mathcal{E}\right\|_2\\\
			&=O_P\left(\frac{1}{T^2} \left(\frac{T}{p_x}+\sqrt{\frac{T}{p_x}}\right)\left(\sqrt{p_x}h_T+1\right)(Th_T+T)\right)\\
			&=O_P\left( \frac{1}{\sqrt{Tp_x}}+\frac{1}{p_x}+\frac{h_T}{\sqrt{p_x}}+\frac{h_T}{\sqrt{T}}\right).
		\end{align*}
		by Lemmas \ref{P-lm.fac.est} \eqref{P-lfaiii}, \ref{lm.H} \eqref{lHi} and \ref{lm.F} \eqref{lfxi}.
		Next, we consider $J_3$. We have 
		\begin{align*}J_3 &\le \frac{1}{T^2}\max_{k \in[p_x]}\left| \sum_{r\in[R]} \left( \mathcal{E}^\top \left(\widehat{F}-FH^\top\right)\right)_{k,r}\left(H F^\top W\right)_{r}\right| \\
			&\le \frac{R}{T^2}\left\|\left(\widehat{F}-FH^\top\right)^\top  \mathcal{E}\right\|_\infty \left\|H F^\top W\right\|_\infty\\
			&\le \frac{R}{T^2}\left\|\left(\widehat{F}-FH^\top\right)^\top  \mathcal{E}\right\|_\infty \max_{r\in[R],k\in[p_x]}\left|\sum_{j\in[R] } H_{r,j} \left(F^\top W\right)_{j,k}\right|\\
			&\le \frac{R^2}{T^2}\left\|\left(\widehat{F}-FH^\top\right)^\top  \mathcal{E}\right\|_\infty\left\|H \right\|_2\left\| F^\top W\right\|_\infty\\\
			&=O_P\left(\frac{1}{T^2} \left(\frac{T}{p_x}+1\right)^{1/2}(Th_T+T)\right)\\
			&=o_P\left(\frac{1}{\sqrt{Tp_x}} +\frac{1}{T}\right),
		\end{align*}
		by Lemmas \ref{P-lm.fac.est} \eqref{P-lfaii}, \ref{lm.H} \eqref{lHi} and \ref{lm.F} \eqref{lfxi}.
		Then, we focus on $J_4$. It holds that 
		\begin{align*}
			J_4&\le\frac{1}{T^2}\max_{k\in[p_x]}\left| \sum_{r\in[R]} \left(W^\top F \right)_{k,r}\left(\left(I_R-H^\top H\right) F^\top  \mathcal{E}\right)_{r}\right|  \\
			&\le \frac{1}{T^2} \|W^\top F\|_\infty \left\|\left(I_R-H^\top H\right) F^\top  \mathcal{E}\right\|_2\\
			&\le \frac{1}{T^2} \|W^\top F\|_\infty \| \mathcal{E}^\top F\|_2\left\| I_R-H^\top H\right\|_2\\
			&= O_P\left(\frac{1}{T^2}  (Th_T+T) \sqrt{T}\left(\frac{1}{\sqrt{T}}+\frac{1}{\sqrt{p_x}}\right)\right)\\
			&= O_P\left(\frac{1}{T}+\frac{1}{\sqrt{Tp_x}}\right),
		\end{align*}
		by Lemmas \ref{lm.F} \eqref{lfxi}, \eqref{lfv} and \ref{lm.H} \eqref{lHii}.
		Next, we bound $J_5$, We have 
		\begin{align*}
			J_5 &\le\frac{1}{T^2} \max_{k\in[p_x]}\left| \sum_{r\in[R]} \left(W^\top F \right)_{k,r}\left( F^\top  \mathcal{E}\right)_{r}\right| \\
			&\le \frac{R}{T^2} \left\| W^\top F\right\|_\infty \left\|F^\top  \mathcal{E}\right\|_\infty\\
			& \le \frac{R}{T^2} \left\| W^\top F\right\|_\infty \left\|F^\top  \mathcal{E}\right\|_2 \\
			&= O_P\left(\frac{1}{T^2}  (Th_T+T) \sqrt{T}\right)= O_P\left(\frac{1}{\sqrt{T}}\right),
		\end{align*}
		Finally, we have
		\begin{align*}
			J_6 &=O_P\left(h_T\right)
		\end{align*}
		by Lemma \ref{lm.F} \eqref{lfxvii}. Combining the bounds on $J_1$ to $J_6$, we obtain \eqref{lTii}.\\
		
		\noindent \textit{Proof of \eqref{lTiii}.} 
		We have 
		\begin{align*}
			\left\|M_{\widehat{F}}F\gamma\right\|_2 = \left\|\left(I_T-\frac1T\widehat{F}\widehat{F}^\top\right)F\gamma\right\|_2\le J_1 + J_2 +J_3+J_4+J_5,\end{align*}
		where 
		\begin{align*}
			J_1&=  \frac1T\left\|\left(\widehat{F}-FH^\top\right) H F^\top F\gamma \right\|_2;\\
			J_2&= \frac1T\left\| FH^\top \left(\widehat{F}-FH^\top\right)^\top F\gamma\right\|_2;\\
			J_3&=  \frac1T\left\|\left(\widehat{F}-FH^\top\right) \left(\widehat{F}-FH^\top\right)^\top F\gamma\right\|_2;\\
			J_4&=  \frac1T\left\|F\left(H^\top H-I_R\right) F\gamma\right\|_2;\\
			J_5&=\left\| \left(I_T - \frac1T F F^\top \right)F\gamma\right\|_2.
		\end{align*}
		We consider first $J_1$. It holds that 
		\begin{align*}J_1&\le  \frac1T  \left\|\widehat{F}-FH^\top\right\|_2 \|F\|_2^2\|H\|_2\|\gamma\|_2\\
			& =O_P\left(\frac1T \left(\frac{T}{p_x}+1\right)^{1/2} T \right)=O_P\left(\sqrt{\frac{T}{p_x}+1}\right),
		\end{align*}
		by Lemmas \ref{P-lm.fac.est} \eqref{P-lfai} and \ref{lm.F} \eqref{lfiii}.
		Similarly, we have $$J_2=O_P\left(\sqrt{\frac{T}{p_x}+1}\right).$$ Next, we focus on $J_3$ and obtain 
		\begin{align*}J_3&\le  \frac1T  \left\|\widehat{F}-FH^\top\right\|_2^2 \|F\|_2\|\gamma\|_2\\
			& =O_P\left(\frac1T \left(\frac{T}{p_x}+1\right)\sqrt{T} \right)=O_P\left(\frac{\sqrt{T}}{p_x}+\frac{1}{\sqrt{T}}\right)=O_P\left(\sqrt{\frac{T}{p_x}+1}\right).,
		\end{align*}
		by Lemmas \ref{P-lm.fac.est} \eqref{P-lfai} and \ref{lm.F} \eqref{lfiii}. Then, we consider $J_4$. We have 
		\begin{align*}
			J_4&\le \frac1T \|F\|_2^2 \|\gamma\|_2 \left\|H^\top H-I_R\right\|_2\\
			&= O_P\left(\frac{1}{\sqrt{T}}+\frac{1}{\sqrt{p_x}}\right),
		\end{align*}
		by Lemmas \ref{lm.F} \eqref{lfiii} and \ref{lm.H} \eqref{lHii}.
		Finally, we have 
		\begin{align*}J_5&=\left\| \left(I_T - \frac1T F F^\top \right)F \left(I_R - \frac1T F^\top F \right)\gamma\right\|_2\\
			&\le   \left\|I_R - \frac1T F^\top F\right\|_2 \|F\|_2\|\gamma\|_2\\
			& =O_P\left(\frac{1}{\sqrt{T}}\sqrt{T} \right)=O_P\left(1\right),\end{align*}
		by Lemma \ref{lm.F} \eqref{lfi} and \eqref{lfiii}. Combining the bounds on $J_1$ to $J_5$, we obtain \eqref{lTiii}.\\
		
		\noindent \textit{Proof of \eqref{lTiv}.} 
		It holds that 
		\begin{align*}
			\left\|P_{\widehat{F}}\mathcal{E}\right\|_2&= \left\|\frac1T\widehat{F}\widehat{F}^\top  \mathcal{E}\right\|_2\\
			&\le \frac1T \left\| \widehat{F} \left(\widehat{F}-FH^\top\right) \mathcal{E}\right\|_2 + \frac1T \left\|\widehat{F}F^\top  \mathcal{E}\right\|_2\\
			& \le \frac1T \left\|\widehat{F}\right\|_2  \left\|\widehat{F}-FH^\top\right\|_2 \| \mathcal{E}\|_2 +\frac1T  \left\|\widehat{F}\right\|_2 \|F^\top  \mathcal{E}\|_2\\
			&=O_P\left( \frac1T\sqrt{T} \left(\frac{T}{p_x}+1\right)^{1/2}\sqrt{T} +\frac1T \sqrt{T}\sqrt{T}\right)\\
			&=O_P\left(\sqrt{\frac{T}{p_x}+1}\right),
		\end{align*}
		by Lemmas \ref{P-lm.fac.est} \eqref{P-lfai} and \ref{lm.F} \eqref{lfiv}, \eqref{lfv} and the fact that $\left\|\widehat{F}\right\|_2=\sqrt{RT}.$
	\end{Proof}

	\subsection{Pre-existing results}\label{sec.pre}
	
	The following lemma is a direct consequence of the Fuk-Nagaev inequality for $\tau$-mixing processes of \cite{babii2022machine} (see their Theorem A.1).
	\begin{Lemma}\label{lm.bab1} Let $\{\zeta_t\}_t$ be a $p$-dimensional mean zero stationary random process such that 
		\begin{enumerate}[\textup{(}i\textup{)}]  
			\item\label{FNi} for some $m>2$, $\vertiii{\zeta_{t}}_m=O(1)$;
			\item\label{FNii} there exists $c>0$ and $a>(m-1)/(m-2)$ such that, for every $j\in[p]$, the $\tau$-mixing coefficients of $\{\zeta_t\}_t$ satisfy $\tau_s^{(j)}\le cs^{-a}$.
		\end{enumerate}
		Then, we have 
		$$\left\| \frac1T \sum_{t\in[T]}\zeta_t  \right\|_\infty=O_P\left(\left(\frac{p}{T^{\kappa-1}}\right)^{1/\kappa}\vee \sqrt{\frac{\log(2p)}{T}}\right),$$
		where $\kappa= \frac{(a+1)m-1}{a+m-1}$. 
	\end{Lemma}
	The following lemma is a direct implication of Lemma A.1.2 in \cite{babii2024high}.
	\begin{Lemma}\label{lm.bab2} Let $\{\zeta_t\}_t$ be a $1$-dimensional mean zero stationary random process such that, for some $q>2$, $\vertiii{\zeta_{t}}_q=O(1)$. Then, we have $$|\E[\zeta_0\zeta_t]|\le \tau_s^{\frac{q-2}{q-1}} \vertiii{\zeta_t}_q^{\frac{q}{q-1}},$$
		where $\{\tau_s\}_s$ are the $\tau$-mixing coefficients of $\{\zeta_t\}_t$.
	\end{Lemma}
	The following Lemma is a direct consequence of Lemma A.2.1 in \cite{babii2022machine}.
	\begin{Lemma}\label{lm.bab3}
		For any $d\in\R^T$, we have 
		$$\Omega^*\left(d\right)=\mu \left\|d\right\|_\infty+(1-\mu)\max_{G\in\mathcal{G}}\left\|d_G\right\|_2.$$
		This implies that 
		$$\Omega^*\left(d\right)\le\left(\mu +(1-\mu)\sqrt{G^*}\right)\left\|d\right\|_\infty.$$
	\end{Lemma}
	
	\section{Monte Carlo results \label{app:mc}}
	\captionsetup{skip=0pt,font={stretch=1.6}}
	\begin{table}
		\centering
		\begin{tabular}{rrrrrrrr}
			& M1 & M2 & M3 &  & M1 & M2 & M3 \\ 
			\hline
			&\multicolumn{3}{c}{Gaussian} && \multicolumn{3}{c}{student-$t(5)$}\\
			\multicolumn{8}{c}{Panel A. {\it Baseline} $(c_z = c_u=0.1)$}\\
			T = 50 & 1.5023 & 1.3753 & 1.1760 &  & 3.2512 & 2.2274 & 1.9531 \\ 
			100 & 1.4983 & 1.2031 & 0.8992 &  & 2.5382 & 1.9978 & 1.5269 \\ 
			200 & 1.3992 & 0.9425 & 0.6133 &  & 2.4736 & 1.6820 & 1.1065 \\ 
			\multicolumn{8}{c}{Panel B. {\it Stronger cross-sectional dependence} $(c_z = c_u=0.4)$}\\
			T = 50 & 1.5686 & 1.4210 & 1.2540 &  & 3.0348 & 2.1126 & 2.0059 \\ 
			100 & 1.4382 & 1.2322 & 0.8883 &  & 2.3524 & 1.9178 & 1.4572 \\ 
			200 & 1.3819 & 0.8924 & 0.5658 &  & 2.2625 & 1.6933 & 1.1337 \\ 
			\hline\hline
		\end{tabular}
		\caption{Monte Carlo  MSE comparisons --- M1 - FAMIDAS, M2 - sg-LASSO-MIDAS, M3 - sg-LASSO-FAMIDAS.\label{tab:mcs}}
	\end{table}
	
	\begin{table}
		\centering
		\begin{tabular}{rrrrrrrr}
			& M1 & M2 & M3 &  & M1 & M2 & M3 \\ 
			\hline
			&\multicolumn{3}{c}{Gaussian} && \multicolumn{3}{c}{student-$t(5)$}\\
			\multicolumn{8}{c}{Panel A. {\it Baseline} $(c_z = c_u=0.1)$}\\
			T=50 & 0.6321 & 0.7212 & 0.5125 &  & 1.0382 & 1.1356 & 1.0286 \\ 
			100 & 0.4604 & 0.5340 & 0.4929 &  & 0.9911 & 1.1060 & 1.0882 \\ 
			200 & 0.4028 & 0.4269 & 0.3899 &  & 0.7230 & 0.9932 & 0.7312 \\ 
			\multicolumn{8}{c}{Panel B. {\it Stronger cross-sectional dependence} $(c_z = c_u=0.4)$}\\
			T = 50 & 1.2686 & 1.4210 & 1.2540 &  & 2.0348 & 2.1126 & 2.0059 \\ 
			100 & 0.8382 & 1.2322 & 0.8883 &  & 1.4524 & 1.9178 & 1.4572 \\ 
			200 & 0.5819 & 0.8924 & 0.5658 &  & 1.1325 & 1.6933 & 1.1337 \\ 
			\hline\hline
		\end{tabular}
		\caption{Dense — Monte Carlo  MSE comparisons --- M1 - FAMIDAS, M2 - sg-LASSO-MIDAS, M3 - sg-LASSO-FAMIDAS.\label{tab:mcs_dense}}
	\end{table}

	\begin{table}
		\centering
		\begin{tabular}{rrrrrrrr}
			& M1 & M2 & M3 &  & M1 & M2 & M3 \\ 
			\hline
			&\multicolumn{3}{c}{Gaussian} && \multicolumn{3}{c}{student-$t(5)$}\\
			\multicolumn{8}{c}{Panel A. {\it Baseline} $(c_z = c_u=0.1)$}\\
			T=50 & 0.9572 & 0.6759 & 0.7027 &  & 1.9403 & 1.3557 & 1.4591 \\ 
			100 & 0.8178 & 0.4348 & 0.4491 &  & 1.3105 & 0.8873 & 0.9077 \\ 
			200 & 0.7628 & 0.2692 & 0.2833 &  & 1.4940 & 0.4529 & 0.4625 \\ 
			\multicolumn{8}{c}{Panel B. {\it Stronger cross-sectional dependence} $(c_z = c_u=0.4)$}\\
			 T=50 & 1.5344 & 1.3159 & 1.1788 &  & 2.8318 & 2.1018 & 1.9745 \\ 
			100 & 1.4397 & 1.0325 & 0.8774 &  & 2.4107 & 1.6360 & 1.5087 \\ 
			200 & 1.3810 & 0.5957 & 0.5765 &  & 2.3810 & 1.5957 & 1.5765 \\ 
			\hline\hline
		\end{tabular}
		\caption{Sparse — Monte Carlo  MSE comparisons --- M1 - FAMIDAS, M2 - sg-LASSO-MIDAS, M3 - sg-LASSO-FAMIDAS.\label{tab:mcs_sparse}}
	\end{table}

	\clearpage
	
	\section{Additional details on the data \label{app:data}}
	
	\renewcommand{\arraystretch}{0.5}
	\paragraph{Monthly macro  data}
	\renewcommand*{\arraystretch}{1.05}
	\begin{longtable}{rlll}
		& Description & Category & T-code \\ 
		\hline\hline
		1 & Real Personal Income & Output and income & 5 \\ 
		2 & Real personal income ex transfer receipts & Output and income & 5 \\ 
		3 & Industrial Production Index & Output and income & 5 \\ 
		4 & IP: Final Products and Nonindustrial Supplies & Output and income & 5 \\ 
		5 & IP: Final Products (Market Group) & Output and income & 5 \\ 
		6 & IP: Consumer Goods & Output and income & 5 \\ 
		7 & IP: Materials & Output and income & 5 \\ 
		8 & IP: Manufacturing (SIC) & Output and income & 5 \\ 
		9 & Capacity Utilization: Manufacturing & Output and income & 5 \\ 
		10 & Civilian Labor Force & Labour market & 5 \\ 
		11 & Civilian Employment & Labour market & 2 \\ 
		12 & Civilian Unemployment Rate & Labour market & 5 \\ 
		13 & Average Duration of Unemployment (Weeks) & Labour market & 5 \\ 
		14 & Civilians Unemployed - Less Than 5 Weeks & Labour market & 2 \\ 
		15 & Civilians Unemployed for 5-14 Weeks & Labour market & 2 \\ 
		16 & Civilians Unemployed - 15 Weeks \& Over & Labour market & 5 \\ 
		17 & Civilians Unemployed for 15-26 Weeks & Labour market & 5 \\ 
		18 & Civilians Unemployed for 27 Weeks and Over & Labour market & 5 \\ 
		19 & Initial Claims & Labour market & 5 \\ 
		20 & All Employees: Total nonfarm & Labour market & 5 \\ 
		21 & All Employees: Goods-Producing Industries & Labour market & 5 \\ 
		22 & All Employees: Mining and Logging: Mining & Labour market & 5 \\ 
		23 & All Employees: Construction & Labour market & 5 \\ 
		24 & All Employees: Manufacturing & Labour market & 5 \\ 
		25 & All Employees: Durable goods & Labour market & 5 \\ 
		26 & All Employees: Nondurable goods & Labour market & 5 \\ 
		27 & All Employees: Service-Providing Industries & Labour market & 5 \\ 
		28 & All Employees: Wholesale Trade & Labour market & 5 \\ 
		29 & All Employees: Retail Trade & Labour market & 5 \\ 
		30 & All Employees: Financial Activities & Labour market & 5 \\ 
		31 & All Employees: Government & Labour market & 5 \\ 
		32 & Avg Weekly Hours : Goods-Producing & Labour market & 5 \\ 
		33 & Avg Weekly Overtime Hours : Manufacturing & Labour market & 5 \\ 
		34 & Avg Weekly Hours: Manufacturing & Labour market & 1 \\ 
		35 & Avg Hourly Earnings: Goods-Producing & Labour market & 2 \\ 
		36 & Avg Hourly Earnings: Construction & Labour market & 1 \\ 
		37 & Avg Hourly Earnings: Manufacturing & Labour market & 4 \\ 
		38 & Housing Starts: Total New Privately Owned & Housing & 4 \\ 
		39 & Housing Starts, Northeast & Housing & 4 \\ 
		40 & Housing Starts, Midwest & Housing & 4 \\ 
		41 & Housing Starts, South & Housing & 4 \\ 
		42 & Housing Starts, West & Housing & 4 \\ 
		43 & New Private Housing Permits, Northeast (SAAR) & Housing & 4 \\ 
		44 & New Private Housing Permits, Midwest (SAAR) & Housing & 4 \\ 
		45 & New Private Housing Permits, South (SAAR) & Housing & 4 \\ 
		46 & New Private Housing Permits, West (SAAR) & Housing & 5 \\ 
		47 & Real Manu. and Trade Industries Sales & Consumption & 5 \\ 
		48 & Retail and Food Services Sales & Consumption & 5 \\ 
		49 & New Orders for Durable Goods & Consumption & 5 \\ 
		50 & New Orders for Nondefense Capital Goods & Consumption & 2 \\ 
		51 & Unfilled Orders for Durable Goods & Consumption & 6 \\ 
		52 & Total Business Inventories & Consumption & 6 \\ 
		53 & Total Business: Inventories to Sales Ratio & Consumption & 5 \\ 
		54 & Consumer Sentiment Index & Consumption & 6 \\ 
		55 & M1 Money Stock & Money and credit & 7 \\ 
		56 & M2 Money Stock & Money and credit & 6 \\ 
		57 & Real M2 Money Stock & Money and credit & 6 \\ 
		58 & Total Reserves of Depository Institutions & Money and credit & 6 \\ 
		59 & Reserves Of Depository Institutions & Money and credit & 2 \\ 
		60 & Commercial and Industrial Loans & Money and credit & 6 \\ 
		61 & Real Estate Loans at All Commercial Banks & Money and credit & 6 \\ 
		62 & Total Nonrevolving Credit & Money and credit & 6 \\ 
		63 & Nonrevolving consumer credit to Personal Income & Money and credit & 6 \\ 
		64 & Consumer Motor Vehicle Loans Outstanding & Money and credit & 6 \\ 
		65 & Total Consumer Loans and Leases Outstanding & Money and credit & 6 \\ 
		66 & Securities in Bank Credit at All Commercial Banks & Money and credit & 6 \\ 
		67 & Crude Oil, spliced WTI and Cushing & Prices & 6 \\ 
		68 & PPI: Metals and metal products: & Prices & 6 \\ 
		69 & CPI : All Items & Prices & 6 \\ 
		70 & CPI : Apparel & Prices & 6 \\ 
		71 & CPI : Transportation & Prices & 6 \\ 
		72 & CPI : Medical Care & Prices & 6 \\ 
		73 & CPI : Commodities & Prices & 2 \\ 
		74 & CPI : Services & Prices & 6 \\ 
		75 & CPI : All Items Less Food & Prices & 6 \\ 
		76 & CPI : All items less medical care & Prices & 6 \\ 
		\hline\hline
		\caption{FRED MD monthly data subset. Definitions of t-codes are available in the primary data source. \\Source: \href{https://research.stlouisfed.org/econ/mccracken/fred-databases/}{https://research.stlouisfed.org/econ/mccracken/fred-databases/}}
		\label{app:macrodata}
	\end{longtable}
	
	\paragraph{Weekly financial data}
	\begin{longtable}{rlll}
		& Description & Category & T-code \\ 
		\hline\hline
		1 & 1-mo. Nonfinancial commercial paper A2P2/AA credit spread & Credit & 1 \\ 
		2 & Moody's Baa corporate bond/10-yr Treasury yield spread & Credit & 1 \\ 
		3 & BofAML High Yield/Moody's Baa corporate bond yield spread & Credit & 1 \\ 
		4 & 30-yr Jumbo/Conforming fixed rate mortgage spread & Credit & 1 \\ 
		5 & 30-yr Conforming Mortgage/10-yr Treasury yield spread & Credit & 1 \\ 
		6 & 10-yr Constant Maturity Treasury yield & Leverage & 2 \\ 
		7 & S\&P 500 Financials/S\&P 500 Price Index (Relative to 2-yr MA) & Leverage & 5 \\ 
		8 & S\&P 500, S\&P 500 mini, NASDAQ 100, NASDAQ mini Open Interest & Leverage & 4 \\ 
		9 & 3-mo. Eurodollar, 10-yr/3-mo. swap, 2-yr and 10-yr Treasury Open Interest & Leverage & 4 \\ 
		10 & 1-mo. Asset-backed/Financial commercial paper spread & Risk & 1 \\ 
		11 & BofAML Home Equity ABS/MBS yield spread & Risk & 1 \\ 
		12 & 3-mo. Financial commercial paper/Treasury bill spread & Risk & 1 \\ 
		13 & Commercial Paper Outstanding & Risk & 3 \\ 
		14 & BofAML 3-5 yr AAA CMBS OAS spread & Risk & 1 \\ 
		15 & 3-mo./1-wk AA Financial commercial paper spread & Risk & 1 \\ 
		16 & Treasury Repo Delivery Fails Rate & Risk & 4 \\ 
		17 & Agency Repo Delivery Failures Rate & Risk & 4 \\ 
		18 & Government Securities Repo Delivery Failures Rate & Risk & 4 \\ 
		19 & Agency MBS Repo Delivery Failures Rate & Risk & 4 \\ 
		20 & 3-mo. Eurodollar spread (LIBID-Treasury) & Risk & 1 \\ 
		21 & On-the-run vs. Off-the-run 10-yr Treasury liquidity premium & Risk & 1 \\ 
		22 & Fed Funds/Overnight Treasury Repo rate spread & Risk & 1 \\ 
		23 & Fed Funds/Overnight Agency Repo rate spread & Risk & 1 \\ 
		24 & Fed Funds/Overnight MBS Repo rate spread & Risk & 1 \\ 
		25 & 3-mo./1-wk Treasury Repo spread & Risk & 1 \\ 
		26 & 10-yr/2-yr Treasury yield spread & Risk & 1 \\ 
		27 & 2-yr/3-mo. Treasury yield spread & Risk & 1 \\ 
		28 & 10-yr Interest Rate Swap/Treasury yield spread & Risk & 1 \\ 
		29 & 2-yr Interest Rate Swap/Treasury yield spread & Risk & 1 \\ 
		30 & 1-yr Interest Rate Swap/1-Year Treasury spread & Risk & 1 \\ 
		31 & 3-mo. LIBOR/CME Term SOFR-Treasury spread & Risk & 1 \\ 
		32 & 1-yr./1-mo. LIBOR/CME Term SOFR spread & Risk & 1 \\ 
		33 & Advanced Foreign Economies Trade-weighted US Dollar Value Index & Risk & 3 \\ 
		34 & CBOE Market Volatility Index VIX & Risk & 1 \\ 
		35 & 1-mo. BofAML Option Volatility Estimate Index & Risk & 1 \\ 
		36 & 3-mo. BofAML Swaption Volatility Estimate Index & Risk & 1 \\
		\hline\hline
		\caption{Financial weekly data set. Sources: Bloomberg \& Haver Analytics. Definitions of t-codes are available on NFCI Chicago Fed website:\\\href{https://www.chicagofed.org/research/data/nfci/current-data}{https://www.chicagofed.org/research/data/nfci/current-data} }
		\label{app:findata}
	\end{longtable}
	
	\paragraph{Factors}
	
	\begin{longtable}{rll}
		Factor & Frequency & Source website  \\\hline\hline
		ADS & Weekly & \href{https://www.philadelphiafed.org/surveys-and-data/real-time-data-research/ads}{https://www.philadelphiafed.org/surveys-and-data/real-time-data-research/ads}\\
		CFNAI & Monthly & \href{https://www.chicagofed.org/research/data/cfnai/historical-data}{https://www.chicagofed.org/research/data/cfnai/historical-data}\\
		NFCI & Weekly & \href{https://research.stlouisfed.org/econ/mccracken/fred-databases/}{https://research.stlouisfed.org/econ/mccracken/fred-databases/}\\
		\hline\hline
		\caption{Factors.}
		\label{app:factors}
	\end{longtable}
	
	\clearpage
	
	\section{Additional empirical results}
	
	\subsection{Significance testing}\label{sec:pvals}
	
	\begin{longtable}{rlll}
		& sg-LASSO-FAMIDAS & sg-LASSO-MIDAS & \\ 
		\hline
		&\multicolumn{3}{c}{Panel A. {\it Full sample}}\\
		sg-LASSO-MIDAS & 0.016$^{**}$ &  &    \\ 
		FAMIDAS               & 0.020$^{**}$ & 0.020$^{**}$ &    \\ 
		&\multicolumn{3}{c}{Panel B. {\it Up to COVID}}\\
		sg-LASSO-MIDAS & 0.560 &  &    \\ 
		FAMIDAS & 0.028$^{**}$ & 0.070$^*$ &    \\ 
		\hline\hline
		\caption{Nowcast comparisons ---  We report the p-values of the average superior predictive ability bootstrap tests (aSPA) of \cite{quaedvlieg2021multi} over all three horizons comparing sg-LASSO-FAMIDAS, sg-LASSO-MIDAS, and FAMIDAS methods. We test the null hypothesis that the average out-of-sample loss over the three horizons is smaller for the methods in the column versus in the row. $^*$ and $^{**}$ indicates 10\% and 5\% significance, respectively. \label{tab:pvals}}
	\end{longtable}
	
	To assess statistical significance, we apply the average superior predictive ability (aSPA) test proposed by \cite{quaedvlieg2021multi}, which allows for a comprehensive evaluation across three nowcasting horizons. However, it is important to interpret these results with caution, as the test has not been specifically validated for the nowcasting methods used in our empirical application. The outcomes of the aSPA tests are presented in Table \ref{tab:pvals}. Results confirm that for the full sample, factor-augmented sparse MIDAS regression significantly outperforms the two alternative approaches at 5\% significance level. 

	\clearpage

	\subsection{Observed factors}\label{sec:obsf}
	
	\begin{longtable}{rccc}
		& 2-month & 1-month & EoQ \\ 
		\hline
		&\multicolumn{3}{c}{Panel A. {\it Full sample}}\\
		ADS & 0.553 & 0.379 & 0.258 \\   
		CFNAI & 0.325 & 0.446 & 0.334 \\   
		NFCI & 0.699 & 0.505 & 0.549 \\ 
		\hline
		&\multicolumn{3}{c}{Panel B. {\it Up to COVID}}\\
		ADS & 0.810 & 0.782 & 0.742 \\   
		CFNAI & 0.800 & 0.814 & 0.732 \\   
		NFCI & 0.895 & 0.932 & 0.924 \\ 
		\hline\hline
		\caption{Nowcast comparisons --- horizons are 2- and 1-month ahead, as well as the end of the quarter (EoQ). We report results for the full sample in Panel (A), and Panel (B) results excluding the COVID pandemic period, while Panel (C) reports results for the COVID pandemic period and beyond. The out-of-sample period starts from 2008 Q1 to 2022 Q2 (Panel A) and from 2008 Q1 to 2019 Q4 (Panel B). The RMSEs are reported relative to the AR(4) model.\label{tab:obsfactors}}
	\end{longtable}
	
	\subsection{Additional robustness checks}\label{sec:robust}
		
	\begin{table}
		\setlength\extrarowheight{-4pt}
		\centering
		\begin{tabular}{rccc}
			& 2-month & 1-month & EoQ \\ 
			\hline
			&\multicolumn{3}{c}{Panel A. {\it Full sample}}\\
			AR(4) & 9.553 & 9.553 & 9.553  \\ 
			sg-LASSO-FAMIDAS & 0.580 & 0.340 & 0.251  \\ 
			PCA-first & 0.748 & 0.558 & 0.463  \\ 
			sWF & 0.548 & 0.770 & 1.001 \\
			sPCA & 0.689 & 0.631 & 0.727 \\
		    sg-LASSO-FAMIDAS (sWF) & 0.580 & 0.503 &  0.612 \\
		    sg-LASSO-FAMIDAS (sPCA) & 0.569 & 0.424 &  0.513 \\
			&\multicolumn{3}{c}{Panel B. {\it Up to COVID}}\\
			AR(4) & 1.934 & 1.934 & 1.934 \\ 
			sg-LASSO-FAMIDAS & 0.842 & 0.827 & 0.756 \\ 
			PCA-first & 0.877 & 0.876 & 0.769 \\ 
			sWF & 0.998 & 1.030 & 1.015 \\ 
			sPCA & 1.059 & 1.052 & 1.041 \\
			sg-LASSO-FAMIDAS (sWF) & 0.928 & 0.874 & 0.770  \\
			sg-LASSO-FAMIDAS (sPCA) & 0.971 & 0.859 & 0.837 \\ 
			\hline\hline
		\end{tabular}
		\caption{Nowcast comparisons --- horizons are 2- and 1-month ahead, as well as the end of the quarter (EoQ). We report results for the full sample in Panel (A), and Panel (B) results excluding the COVID pandemic period. We report RMSEs for sg-LASSO-FAMIDAS where we i) estimate macro factors using monthly series before applying MIDAS (PCA-first), ii) use sparsity-induced weak factor (sWF) model approach of \cite{uematsu2022estimation}, and use sWF and sparse idiosyncratic components (sg-LASSO-sFAMIDAS (sWF)), and lastly apply sparse PCA of \cite{zou2006sparse} to estimate the factors for factor-only model (sPCA) as well as integrated model of sPCA with sparse idiosyncratic components (sg-LASSO-sFAMIDAS (sPCA)). In all cases, the eigenvalue growth estimator of \cite{ahn2013eigenvalue} is used to determine the number of factors. The out-of-sample period starts from 2008 Q1 to 2022 Q2 (Panel A) and from 2008 Q1 to 2019 Q4 (Panel B). The RMSEs are reported relative to the AR(4) model. \label{tab:pca_diff}}
	\end{table}
	
	\begin{table}
		\setlength\extrarowheight{-4pt}
		\centering
		\begin{tabular}{rccc}
			& 2-month & 1-month & EoQ \\ 
			\hline
			&\multicolumn{3}{c}{Panel A. {\it Full sample}}\\
			AR(4)& 9.553 & 9.553 & 9.553 \\ 
			IC & 0.876 & 0.621 & 0.398 \\ 
			PC1 & 0.876 & 0.621 & 0.398 \\ 
			PC2 & 0.580 & 0.340 & 0.251 \\ 
			RR & 0.841 & 0.581 & 0.331 \\ 
			EV & 0.580 & 0.340 & 0.251 \\
			&\multicolumn{3}{c}{Panel B. {\it Up to COVID}}\\
			AR(4) & 1.934 & 1.934 & 1.934 \\ 
			IC	& 1.554 & 1.182 & 1.537 \\ 
			PC1	 & 1.554 & 1.182 & 1.537 \\ 
			PC2	& 0.842 & 0.827 & 0.756 \\ 
			RR    & 0.892 & 1.108 & 0.901 \\ 
			EV    & 0.842 & 0.827 & 0.756 \\
			\hline\hline
		\end{tabular}
		\caption{Nowcast comparisons --- horizons are 2- and 1-month ahead, as well as the end of the quarter (EoQ). We report results for the full sample in Panel (A), and Panel (B) results excluding the COVID pandemic period. The out-of-sample period starts from 2008 Q1 to 2022 Q2 (Panel A) and from 2008 Q1 to 2019 Q4 (Panel B). The RMSEs are reported relative to the AR(4° model. Relative to \eqref{P-tab:rmse}, this table shows RMSEs using different estimators for the number of factors. Specifically, IC, PC1 and PC2 are the estimators of \cite{bai2002determining}, while RR is the rank-regularized estimator of \cite{bai2019rank}. Lastly, we report results for the eigenvalue ratio estimator of \cite{ahn2013eigenvalue}, denoted as EV.  \label{tab:bn}}
	\end{table}

	\begin{table}
		\setlength\extrarowheight{-4pt}
		\centering
		\begin{tabular}{rccc}
			& 2-month & 1-month & EoQ \\ 
			\hline
			&\multicolumn{3}{c}{Panel A. {\it Full sample}}\\
			AR(4) & 10.319 & 10.319 & 10.319 \\ 
			FAMIDAS & 0.816 & 0.480 & 0.644 \\ 
			sg-LASSO-MIDAS & 0.610 &  0.447 & 0.568\\
			sg-LASSO-FAMIDAS & 0.548 & 0.264 & 0.251  \\ 
			&\multicolumn{3}{c}{Panel B. {\it Up to COVID}}\\
			AR(4) & 1.944 & 1.944 & 1.944  \\ 
			FAMIDAS & 0.944 & 0.874  & 0.804\\ 
			sg-LASSO-MIDAS &  0.839 & 0.810 & 0.787 \\ 
			sg-LASSO-FAMIDAS & 0.857 & 0.930 & 0.734 \\ 
			\hline\hline
		\end{tabular}
		\caption{Nowcast comparisons --- horizons are 2- and 1-month ahead, as well as the end of the quarter (EoQ). We report results for the full sample in Panel (A), and Panel (B) results excluding the COVID pandemic period. The out-of-sample period starts from 2008 Q1 to 2022 Q2 (Panel A) and from 2008 Q1 to 2019 Q4 (Panel B). The RMSEs are reported relative to the AR model. Relative to \eqref{P-tab:rmse}, this table shows RMSEs using a rolling window scheme rather than expanding. The effective sample size is 95 observations, matching the sample size used for the first-quarter nowcasts in our main results. \label{tab:rolling}}
	\end{table}

	\clearpage
	
	\begin{table}
		\setlength\extrarowheight{-4pt}
		\centering
		\begin{tabular}{rccc}
			& 2-month & 1-month & EoQ \\ 
			\hline
			&\multicolumn{3}{c}{Panel A. {\it Full sample}}\\
			AR(4) & 9.553 & 9.553 & 9.553  \\ 
			Macro & 0.580 & 0.340 & 0.251  \\ 
			Financial + macro & 0.611 & 0.315 & 0.356 \\ 
			&\multicolumn{3}{c}{Panel B. {\it Up to COVID}}\\
			AR(4) & 1.934 & 1.934 & 1.934  \\ 
			Macro & 0.842 & 0.827 & 0.756 \\ 
			Financial + macro & 0.958 & 0.883 & 0.818 \\ 
			\hline\hline
		\end{tabular}
		\caption{Nowcast comparisons --- horizons are 2- and 1-month ahead, as well as the end of the quarter (EoQ). We report results for the full sample in Panel (A), and Panel (B) results excluding the COVID pandemic period. The out-of-sample period starts from 2008 Q1 to 2022 Q2 (Panel A) and from 2008 Q1 to 2019 Q4 (Panel B). The RMSEs are reported relative to the AR model. Relative to \eqref{P-tab:rmse}, this table shows results for the sg-LASSO-FAMIDAS approach with two sparse plus dense structures: i) called {\it Macro} (see Table \eqref{P-tab:rmse} rows {\it sg-LASSO-FAMIDAS}, ii) called {\it Financial + macro}, where the regression is the same as {\it Macro} except that it contains an additional set of factors estimated by prinicipal components analysis using the MIDAS-weighted data from the financial panel (the growth ratio estimator is used to determine the number of financial factors). \label{tab:finspd}}
	\end{table}

	\clearpage

	\begin{figure}
		\begin{subfigure}[b]{0.32\linewidth}
			\centering
			\includegraphics[width=1.0\columnwidth]{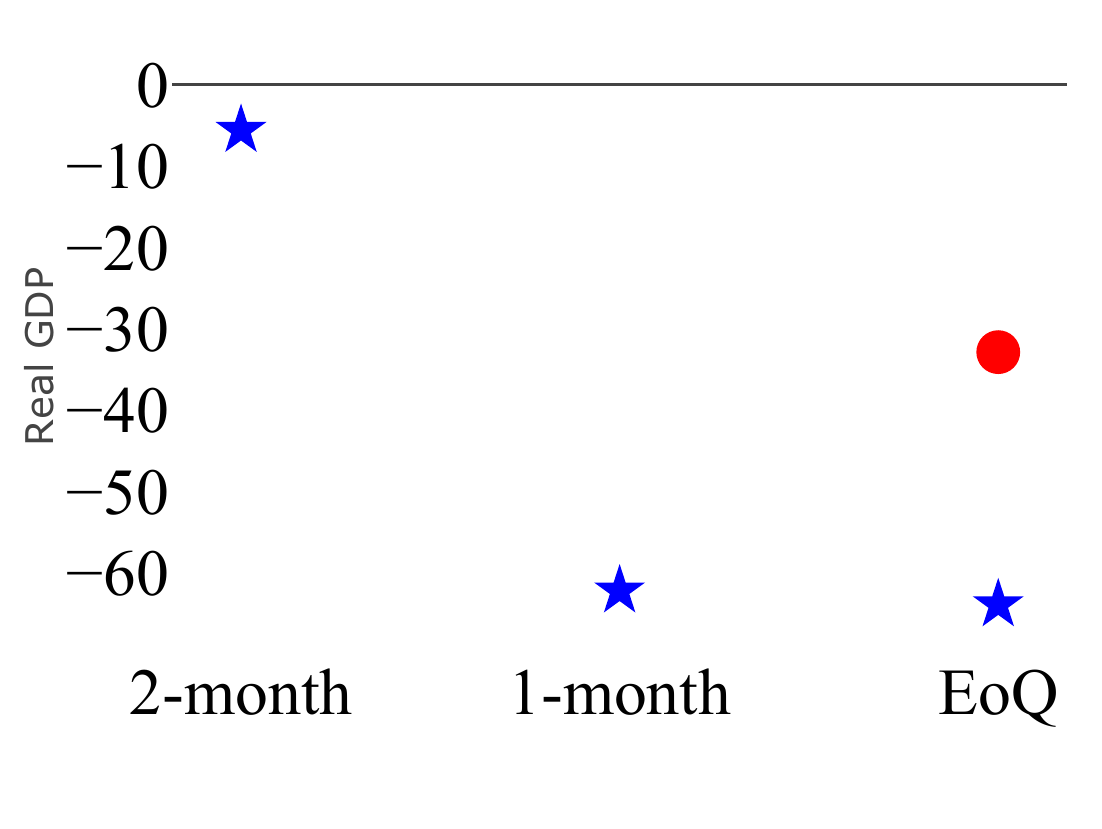}
			\caption{FAMIDAS—Q2}
			\label{fig:nowfamidas}
		\end{subfigure}
		\begin{subfigure}[b]{0.32\linewidth}
			\centering
			\includegraphics[width=1.0\columnwidth]{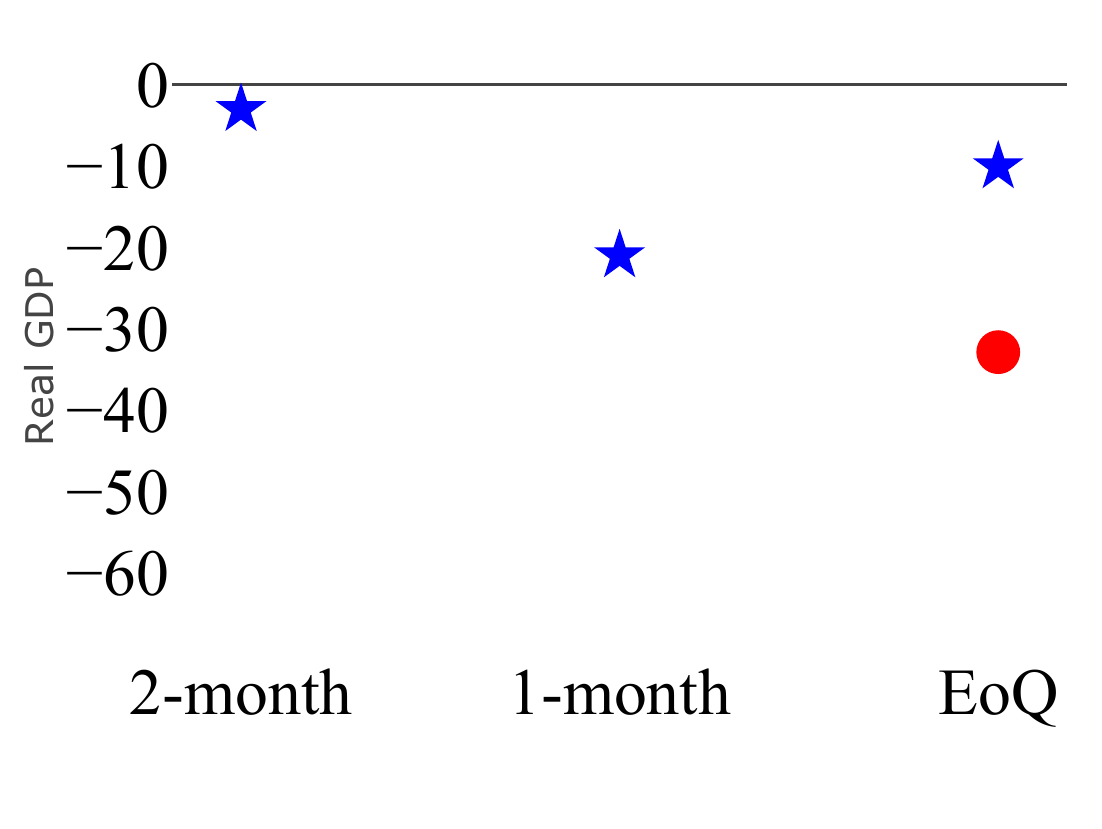}
			\caption{sg-LASSO-MIDAS—Q2}
			\label{fig:nowsglmidas}
		\end{subfigure}
		\begin{subfigure}[b]{0.32\linewidth}
			\centering
			\includegraphics[width=1.0\columnwidth]{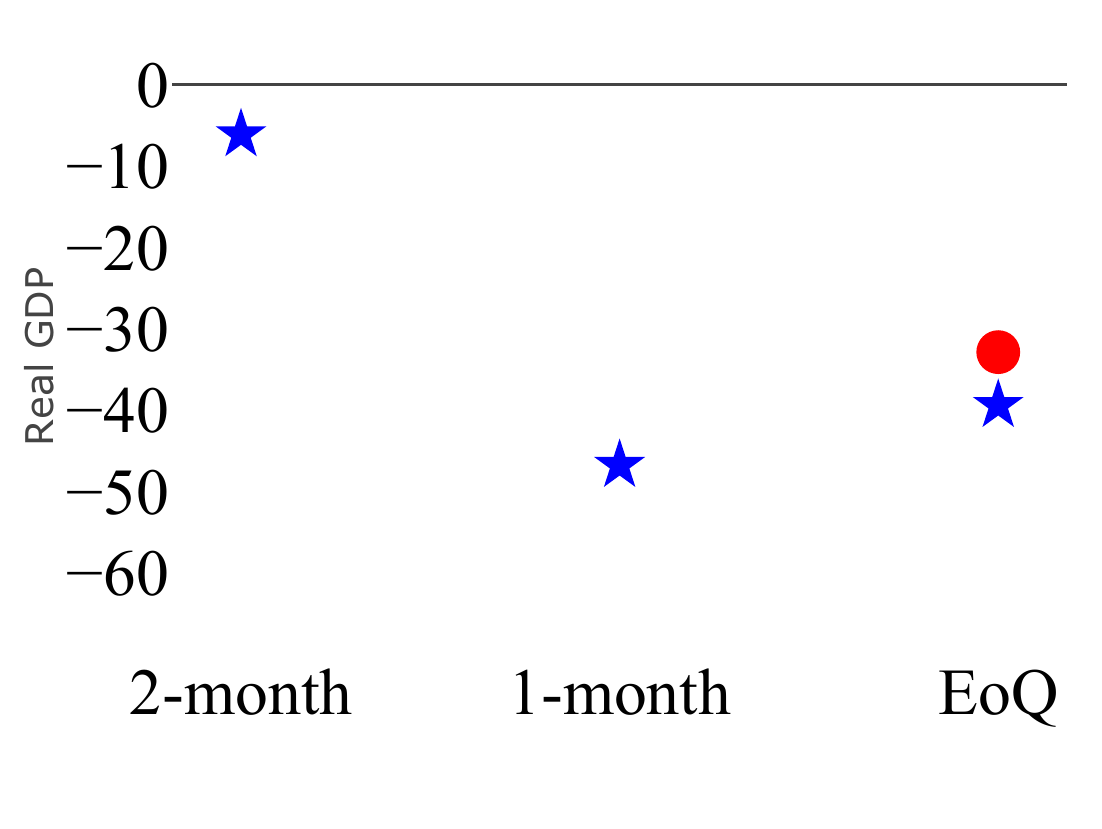}
			\caption{sg-LASSO-FAMIDAS—Q2}
			\label{fig:nowsglfamidas}
		\end{subfigure}\hfill
		\begin{subfigure}[b]{0.32\linewidth}
			\centering
			\includegraphics[width=1.0\columnwidth]{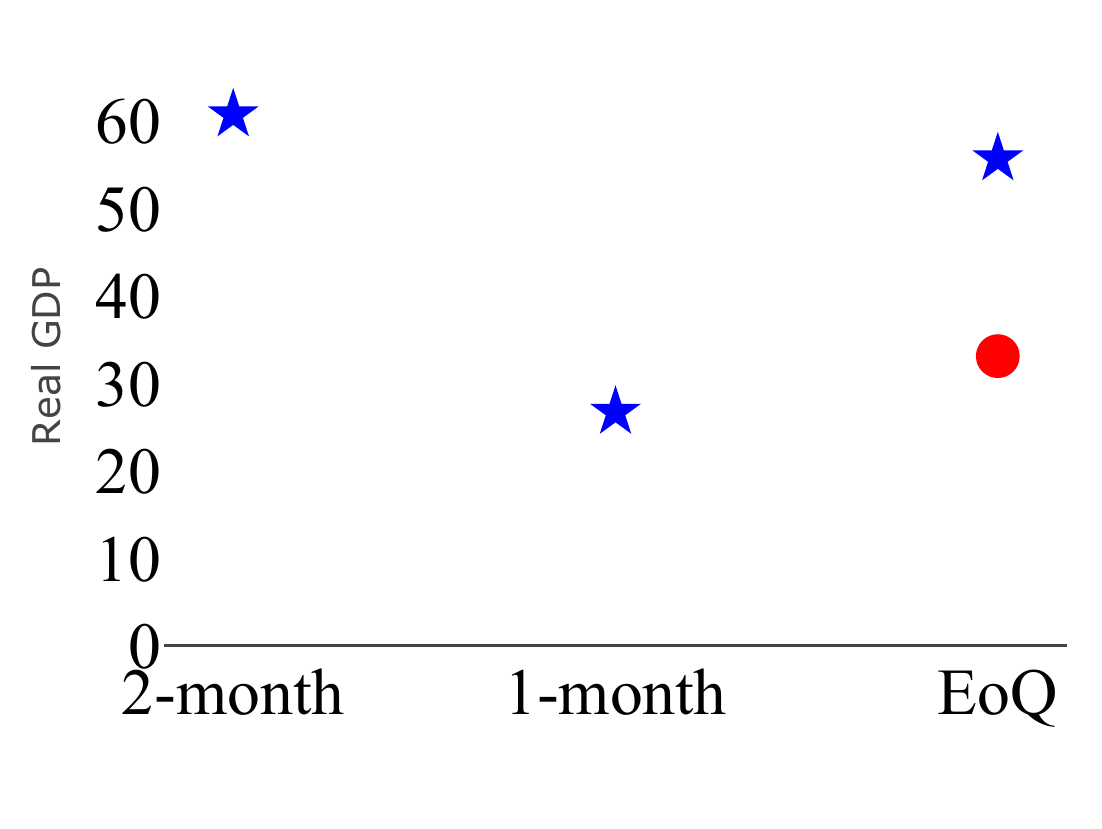}
			\caption{FAMIDAS—Q3}
			\label{fig:nowfamidas_q3}
		\end{subfigure}
		\begin{subfigure}[b]{0.32\linewidth}
			\centering
			\includegraphics[width=1.0\columnwidth]{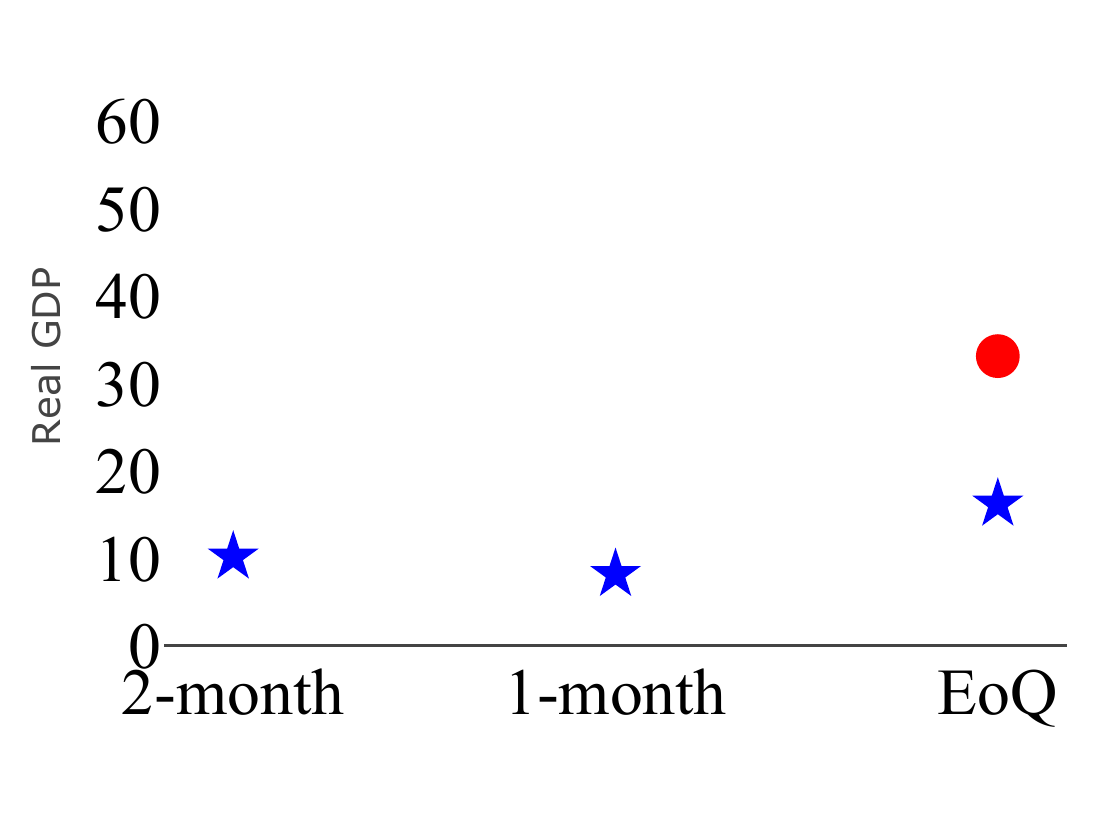}
			\caption{sg-LASSO-MIDAS—Q3}
			\label{fig:nowsglmidas_q3}
		\end{subfigure}
		\begin{subfigure}[b]{0.32\linewidth}
			\centering
			\includegraphics[width=1.0\columnwidth]{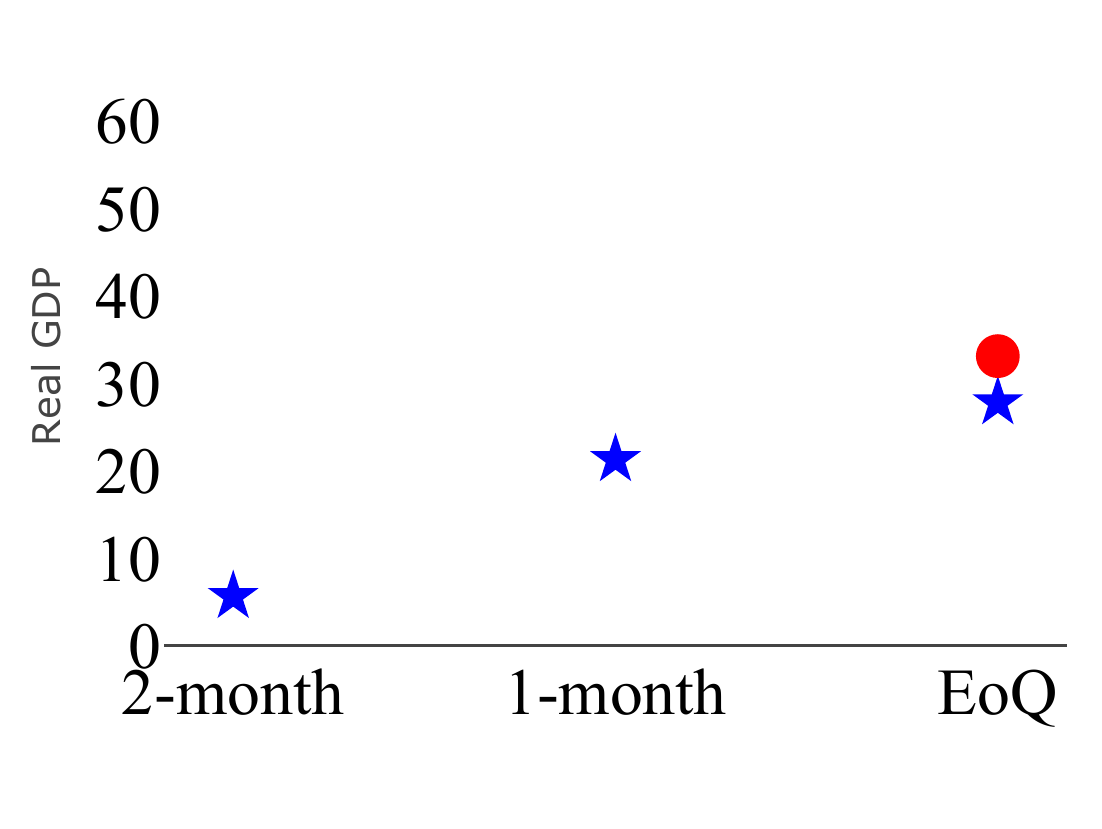}
			\caption{sg-LASSO-FAMIDAS—Q3}
			\label{fig:nowsglfamidas_q3}
		\end{subfigure}\hfill
		\caption{The figure plots the nowcasts of U.S. GDP growth for the FAMIDAS, sg-LASSO-MIDAS and sg-LASSO-FAMIDAS during the pandemic period for the three horizons we consider: 2-month, 1-month and end of quarter (EoQ). The blue star in Figure \ref{fig:nowfamidas}-\ref{fig:nowsglfamidas} plots the nowcasts for the three methods for 2020 Q2 while Figure \ref{fig:nowfamidas_q3}-\ref{fig:nowsglfamidas_q3} plots nowcasts for the same methods but for 2020 Q3. The red circle plots the \textit{advanced release} of the respective quarter, i.e., the target nowcast. \label{fig:now}}
	\end{figure}

	\section{Details on matrix completion}\label{sec.matrixc}
	
	To implement matrix completion, we use the R package \texttt{softImpute}  version 1.4—1 downloaded from CRAN.  The algorithm fits a low-rank matrix approximation to a matrix with missing values via nuclear-norm regularization. We set the maximum number of rank, \texttt{max.rank}, to 6, which restricts the rank of the solution. Starting from $\lambda_0$, where $\lambda$ is the regularization parameter for the nuclear norm minimization problem, we find $\lambda$ so that the solution reached has rank slightly less than \texttt{rank.max}, as suggested in the package manual. $\lambda_0$ is the initial guess, which we set to a value computed by a function \texttt{lambda0} within the package. This function computes the smallest value for $\lambda$ such that \texttt{softImpute} returns the zero solution.

	\bibliographystyle{agsm}
	\bibliography{paper}